\documentclass[fleqn,usenatbib]{mnras}
\usepackage{comment}
\usepackage{setspace}
\usepackage{braket}
\usepackage{url}
\usepackage{longtable}
\usepackage{pdflscape}
\usepackage{bm}
\usepackage{ulem}

\usepackage{newtxtext,newtxmath}

\usepackage[T1]{fontenc}
\usepackage{ae,aecompl}

\usepackage{graphicx}	
\usepackage{amsmath}	

\title[ Cloud-cloud collision speed in NGC~1300]{
Connection among environment, cloud-cloud collision speed, and star formation activity in the strongly barred galaxy NGC~1300}

\author[F. Maeda et al.]{
Fumiya Maeda,$^{1}$\thanks{E-mail: fmaeda@kusastro.kyoto-u.ac.jp}
Kouji Ohta,$^{1}$
Yusuke Fujimoto,$^{2}$
Asao Habe$^{3}$
\\
$^{1}$Department of Astronomy, Kyoto University, Kitashirakawa-Oiwake-Cho, Sakyo-ku, Kyoto, Kyoto 606-8502, Japan\\
$^{2}$Earth and Planets Laboratory, Carnegie Institution for Science, 5241 Broad Branch Road, NW, Washington, DC 20015, USA\\
$^{3}$Graduate School of Science, Hokkaido University, Kita 10 Nishi 8, Kita-ku, Sapporo, Hokkaido 060-0810, Japan
}

\date{Accepted XXX. Received YYY; in original form ZZZ}

\pubyear{2020}

\begin{document}
\label{firstpage}
\pagerange{\pageref{firstpage}--\pageref{lastpage}}
\maketitle

\begin{abstract}

Cloud–cloud collision (CCC) has been suggested as a mechanism to induce massive star formation. Recent simulations suggest that a CCC speed is different among galactic-scale environments, which is responsible for  observed differences in star formation activity. In particular,  a high-speed CCC is proposed as a cause of star formation suppression in the bar regions in barred spiral galaxies. Focusing on the strongly barred galaxy NGC 1300, we investigate the CCC speed.  We find the CCC speed in the bar and bar-end  tend to be higher than that in the arm.  The estimated CCC speed is $\sim20~\rm km~s^{-1}$, $\sim16~\rm km~s^{-1}$, and $\sim11~\rm km~s^{-1}$ in the bar, bar-end, and arm, respectively. Although the star formation activity is different in the bar and bar-end, the CCC speed and the number density of high-speed CCC with $> 20~\rm km~s^{-1}$ are high in both regions, implying the existence of other parameters  that control the star formation.  The difference in molecular gas mass (average density) of the giant molecular clouds (GMCs) between the bar (lower mass and lower density) and bar-end (higher mass and higher density) may be cause for the different star formation activity. Combining with our previous study (Maeda et al.), the leading candidates of causes for the star formation suppression in the bar in NGC~1300 are the presence of a large amount of diffuse molecular gases and high-speed CCCs between low mass GMCs.

\end{abstract}



\defcitealias{fujimoto_fast_2020}{F20}

\section{Introduction} \label{sec: Intro}

Star formation activity changes among galactic-scale environments. Observations of nearby disc galaxies have revealed a large scatter in star formation efficiency (${\rm SFE}= \Sigma_{\rm SFR} / \Sigma_{\rm H_2}$) on kpc/sub-kpc scale \citep[e.g.][]{leroy_molecular_2013}. In particular, the SFE strongly depends on galactic structure such as spiral arms, bar, and nucleus; the SFE in bar regions tends to be lower than that in other regions
\citep[e.g.][]{momose_star_2010,hirota_wide-field_2014, maeda_large_2020}.
Such suppression of massive star formation in the bar regions was reported by other observational studies \citep[e.g.][]{downes_co_1996, james_h_2009,hakobyan_supernovae_2016}. Therefore, in order to understand the diversity of galactic-scale SFE, it is important to understand physical mechanisms that suppress the massive star formation in the bar regions.



Some explanations have been proposed as the cause for the star formation suppression in the bar regions; the strong shock or/and shear \citep[e.g.][]{tubbs_inhibition_1982,athanassoula_existence_1992,reynaud_kinematics_1998, emsellem_interplay_2015}, gas depletion due to the formation of the bar structure \citep[e.g.][]{spinoso_bar-driven_2017,james_star_2018}, the presence of a large amount of diffuse molecular gas \citep[e.g.][]{muraoka_co_2016,yajima_co_2019, torii_forest_2019, maeda_large_2020}, and the presence of gravitationally  unbound giant molecular clouds  \citep[GMCs;][]{sorai_properties_2012,nimori_dense_2013}. However, the physical mechanisms are still unclear.

As another candidate of the cause for the suppression, it is proposed that high-speed cloud-cloud collisions (CCCs) occur in the bar regions.
The concept of CCCs had been previously proposed before \citep[e.g.][]{stone_collisions_1970-1,stone_collisions_1970,loren_colliding_1976,gilden_clump_1984,scoville_high-mass_1986}, and the CCC models as a triggering mechanism of massive star formation were proposed \citep[e.g.][]{Habe1992PASJ, Fukui2014ApJ...780}; CCCs induce clump formation by shock compressions at the collision interface. \citet{tan_star_2000} proposed a model in which the majority of star formation in disc galaxies is induced by CCCs. After that, the role of CCCs in the star formation activity of the galaxies and the impact of the galactic structure on the CCCs have been investigated by three-dimensional hydrodynamical simulations of disc galaxies \citep[e.g.][]{tasker_star_2009,fujimoto_giant_2014, dobbs_frequency_2015,renaud_environmental_2015}.

Recent simulation results of CCCs suggest that star formation activity strongly depends on the relative speed of the two colliding clouds. \citet{takahira_cloud-cloud_2014, takahira_formation_2018}
performed sub-parsec scale simulations of CCCs of two clouds with different CCC speeds of $3 - 30~\rm km~s^{-1}$ and found a faster CCC can shorten the gas accretion phase of the cloud cores formed, leading to suppression of core growth and massive star formation. A high-resolution ($\sim$ a few pc) three-dimensional hydrodynamical simulation of an intermediate-type barred galaxy performed by \citet{fujimoto_giant_2014} found that CCC speed in the bar regions is larger than in the arm regions. Based on this simulation,  \citet{fujimoto_environmental_2014} suggested that the observed low SFE in the bar in nearby galaxies can be explained quantitatively by fast CCCs with assumptions that collisions between $10$ and $40~\rm km~s^{-1}$ are effective to form stars while CCCs with $< 10~\rm km~s^{-1}$ and $>40~\rm km~s^{-1}$ are inefficient. Although the range of CCC speed used in their assumption is somewhat rough (in fact, \citealt{takahira_cloud-cloud_2014,takahira_formation_2018} showed that the efficiency of massive core formation is anticorrelated with the CCC speed in the range of $>10~\rm km~s^{-1}$), they demonstrated that differences in CCC speed among the environments would be responsible for the observed differences in SFE.  According to these results, to understand the diversity of galactic-scale SFE, 
the connection among the environment, the CCC speed, and the star formation activity should be investigated observationally.

In the Milky Way, the number of CCC samples has been increasing recently \citep[more than 80; see ][]{fukui_cloud-cloud_2020}, and the dependence of the peak column density of colliding clouds and relative velocity of them on the number of OB stars is revealing \citep[][]{enokiya_cloud-cloud_2019}; OB stars tend to reside in colliding molecular clouds with higher peak column density and large relative velocity. On the other hand, a high-speed CCC without OB stars is observed in the central region of the Milky Way, which supports the conclusion by \citet{takahira_cloud-cloud_2014}. However, CCC speed-environment connection in nearby galaxies is still unclear. Recent measurements of Oort parameters in disc galaxies suggest that star formation rates at a few hundred-parsec scale are driven by CCCs \citep{aouad_coupling_2020}.
Although direct observations of extragalactic CCCs have been made (e.g. LMC; \citealt{Fukui_2015ApJ...807L...4F}, \citealt{Saigo_2017ApJ...835..108S}, M33; \citealt{sano_alma_2020, tokuda_alma_2020, muraoka_alma_2020}, Antennae galaxies; \citealt{Finn_2019ApJ...874..120F}),  the sample is still not large enough to investigate the connection, and the CCC in the bar regions of nearby galaxies has not been observed within our knowledge.

To unveil the cause for the star formation suppression, we have investigated the properties of molecular gases in strongly barred galaxies as ideal laboratories \citep{maeda_large_2018,maeda_properties_2020,maeda_large_2020,fujimoto_fast_2020}. While  H\textsc{ii} regions  (i.e. massive star formation) are seen in arm regions, prominent H\textsc{ii} regions are mostly not seen in the strong bar  despite the presence of  molecular gases \citep[e.g. NGC~1300 and NGC~5383;][]{tubbs_inhibition_1982,maeda_large_2018}.  In particular, the diversity of SFE on the galactic scale is quantitatively demonstrated in NGC 1300 \citep{maeda_large_2020}; SFE is $0.08-0.12~\rm Gyr^{-1}$, $0.61-0.77~\rm Gyr^{-1}$, and $0.45-0.53~\rm Gyr^{-1}$ in the bar, arm, and bar-end regions, respectively.  This result indicates that the star formation in the bar regions of the strongly barred galaxies is further suppressed than in the intermediate-type barred galaxies; SFE difference between the arm and bar regions in the intermediate-type is a factor of $\sim 2-3$ (e.g. M83; \citealt{hirota_wide-field_2014}, NGC~4303; \citealt{momose_star_2010}). In such strongly barred galaxies, the physical mechanism(s) of the star formation suppression in the bar regions is(are) expected to be clearly observed.

\citet{maeda_large_2020} found that the SFE decreases with increasing the ratio of the diffuse molecular gas surface density to the total molecular gas surface density. This result supports the possibility that the presence of a large amount of diffuse molecular gas, which would not contribute to the star formation, makes the SFE low in appearance. Furthermore, they pointed out that there is another mechanism that suppresses star formation in the GMCs themselves and suggested that the high-speed CCC is a candidate. \citet[\citetalias{fujimoto_fast_2020}]{fujimoto_fast_2020} presented a hydrodynamical simulation of a strongly barred galaxy using a stellar potential model of NGC~1300 and showed that the CCC speed in the bar regions is significantly faster than those in the arm and bar-end regions due to the highly elongated gas motion by the bar potential. They concluded that high-speed CCCs would be one of the physical mechanisms that make the star formation activity extremely low in the strong bar. \citet{maeda_large_2018} carried out CO observations with a single dish with a single-dish telescope (beam size of $1-2$ kpc)  towards NGC~1300 and NGC~5383 and showed the velocity width of the CO line profile tends to be larger in the bar regions than in the arm regions, which is qualitatively consistent with the scenario that the high-speed CCCs suppress the massive star formation in the bar regions. However, the difference in the observed velocity width is not so large. Further, the velocity width is also affected by the velocity field and molecular cloud distribution in the beam.  Therefore, they warned that it was premature to conclude that the results support the high-speed CCCs scenario, and they emphasized that direct GMC observations at a high angular resolution are indispensable to examine the scenario.

\begin{figure}
	\includegraphics[width=\hsize]{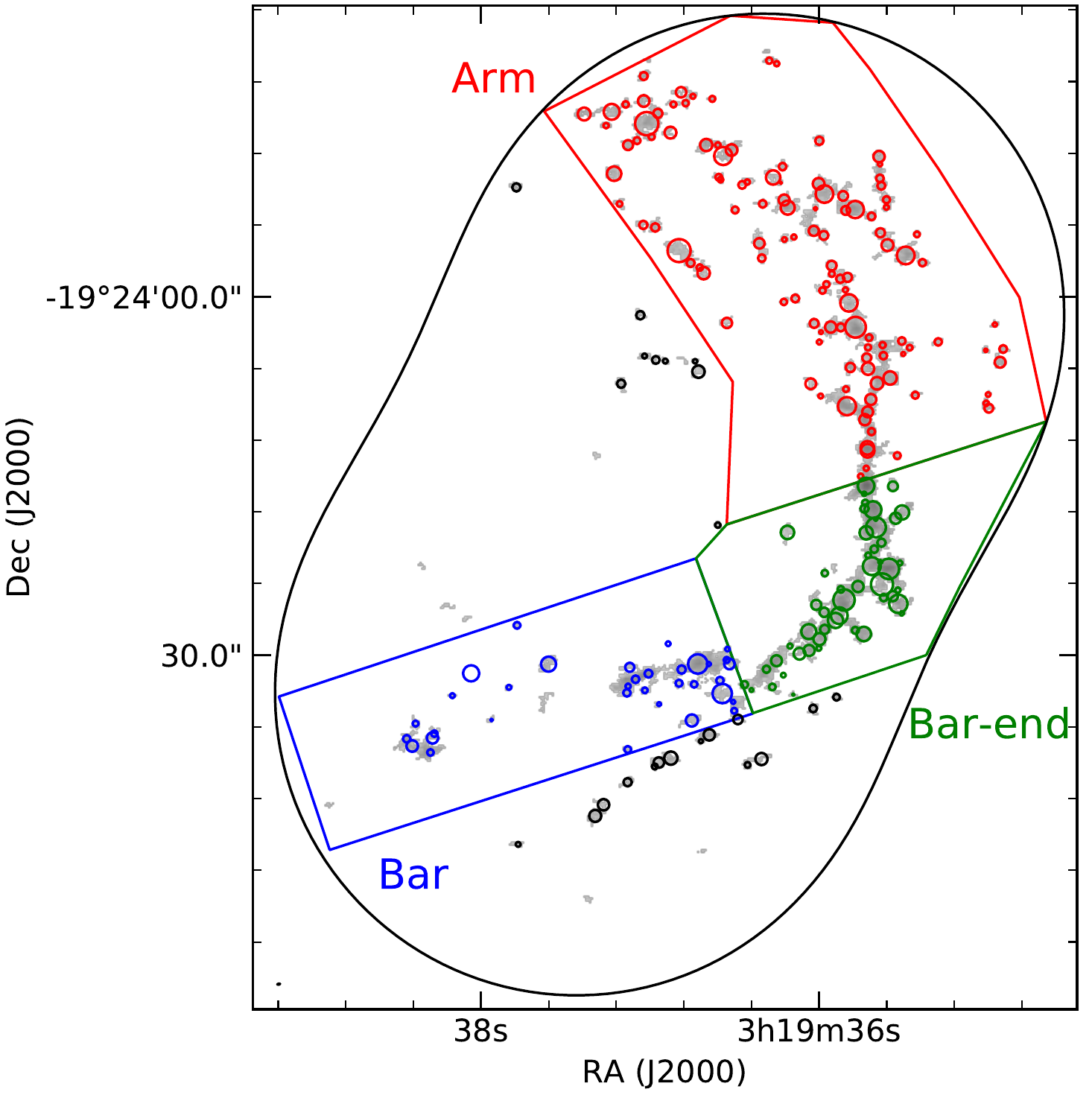}
    \caption{GMC distribution in NGC~1300 superimposed on the velocity-integrated CO($1-0$) map (grey-scale). The GMCs are represented as circles with measured radius, $R$. The colour solid lines indicate the definition of the environmental mask defined by \citet{maeda_properties_2020}. {\it Bar}, {\it arm}, and {\it bar-end} are indicated with blue, red, and green lines, respectively.   The black line represents the FoV of CO($1-0$) observations with ALMA.}
 \label{fig:GMC_positions}
\end{figure}

In this paper, we investigate the CCC speed and its possible effect on the SFE in the strongly barred galaxy NGC~1300 using  $^{12}$CO($1-0$) observations at a high angular resolution of $\sim$40 pc with Atacama Large Millimeter/submillimeter Array (ALMA), which detected more than 200 GMCs \citep{maeda_properties_2020}. Direct observation of CCC in NGC~1300 is desired, but it is difficult with the current angular resolution of our (and archive) data. Therefore, we attempt to estimate the CCC speed of GMCs by predicting the motion of the GMCs based on the mean velocity field of the molecular gas and catalogued line-of-sight velocity of the GMCs. Besides this method, we  measure velocity deviation between the GMC and its surrounding GMCs, which is a model-independent observable and  a good indicator of the qualitative trends of the CCC speed. We have two main goals: first, we examine the differences of the estimated CCC speed and the velocity deviation between the environments (i.e. bar, arm, bar-end; Fig.~\ref{fig:GMC_positions}).
The second goal is to compare the results with the simulation by \citetalias{fujimoto_fast_2020}.
This paper is organized as follows: In Section~\ref{sec: GMC catalogues}, we describe the GMC catalogue we use. In Section~\ref{sec: Estimated CCC speed}, we estimate the CCC speeds using the catalogue and a modeled mean velocity field of the molecular gas. Then we investigate their differences among the environments and compare the observation with the simulation. Section~\ref{sec: velocity deviation} shows the results of the velocity deviation and comparison with the simulation. In Section~\ref{sec:Discussion}, we comprehensively discuss the physical mechanisms controlling galaxy-scale SFE diversity, including other ideas besides the scenario based on CCC speed differences.
Table~\ref{tab:NGC1300} summarizes parameters of NGC~1300 adopted throughout this paper.

\begin{table}
 \caption{Adopted parameters of NGC~1300}
 \label{tab:NGC1300}
 \begin{tabular}{lc}
  \hline
  Parameter & Value \\
  \hline
  Morphology$^a$ & SB(s)bc \\
  Centre position (J2000.0)$^b$ & $\rm 03^h19^m41^s.11$  \\
  & $\rm -19^\circ24^\prime37^{\prime\prime}.7$\\
  Inclination$^c$ & $50^\circ.2$ \\
  PA of the line of node$^c$ & $274^\circ.5$ \\
  PA of the bar$^c$ & $281^\circ$ \\
  Distance$^d$ & 20.7 Mpc \\
  Linear scale & 100 $\rm pc~arcsec^{-1}$ \\
\hline
\multicolumn{2}{l}{{\small$^a$  \citet{Sandae_Tammann}}} \\
\multicolumn{2}{l}{{\small $^b$  The peak of velocity-integrated intensities map of  }}\\
\multicolumn{2}{l}{{\small 
CO($2-1$) line (this work).}}\\
\multicolumn{2}{l}{{\small$^c$  \citet{england_high-resolution_1989}}}\\
\multicolumn{2}{l}{{\small$^d$ We adopted the systemic velocity with corrections for }}\\
\multicolumn{2}{l}{{\small  the Virgo cluster, the Great Attractor, and the Shapley}}\\
\multicolumn{2}{l}{{\small  concentration of $1511~{\rm km~s^{-1}}$ \citep{MouldEtAl00} and }}\\
\multicolumn{2}{l}{{\small  the Hubble constant of $73~{\rm km~s^{-1}~Mpc^{-1}}$. }}
 \end{tabular}
\end{table}

\begin{figure*}
	\includegraphics[width=\hsize]{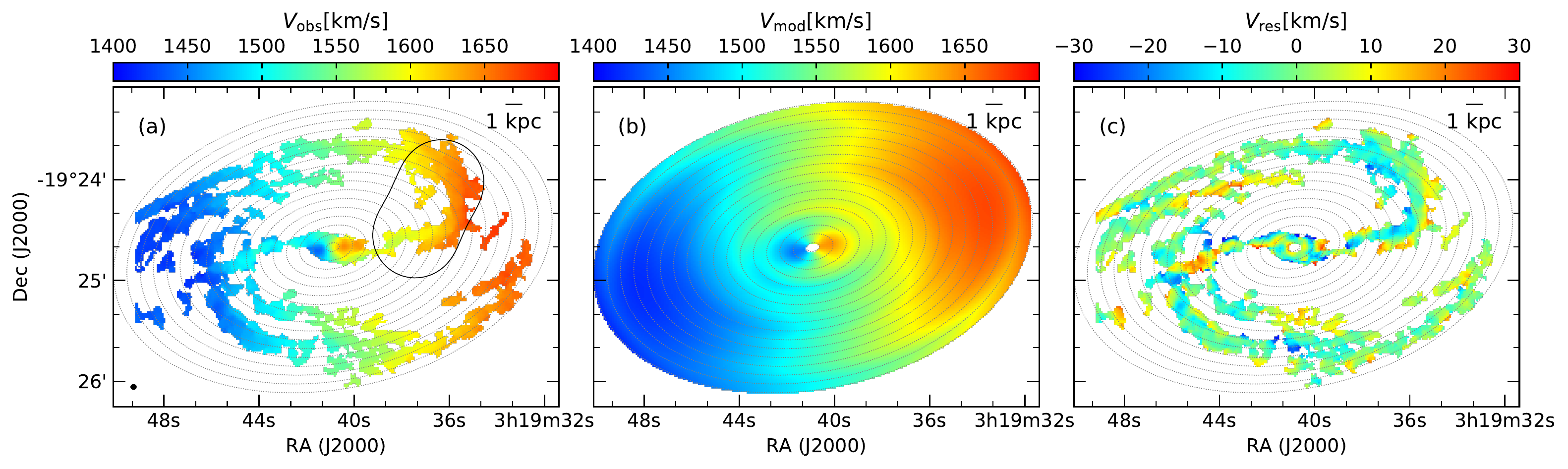}
    \caption{Kinematic models of NGC~1300 velocity field. (a) Observed velocity field obtained from CO($2-1$) observations with ALMA. The black line represents the FoV of CO($1-0$) observations with ALMA. The black circle of the left bottom corner represents the beam-size of 4.0 arcsec. (b) Best-fitted modeled velocity field obtained from \textsc{diskfit}. (c) Residual map.
    We project the model and fit to the observed data, using linear interpolation to predict data values between rings. 
    The radii of the rings set from 0.4 kpc to 13.2 kpc in a 0.8-kpc interval shown as gray dotted lines.}
 \label{fig:Diskfit_obs_model_res_fits}
\end{figure*}

\section{GMC catalogue} \label{sec: GMC catalogues}
GMC catalogue to be used in this study was created by \citet{maeda_properties_2020}. The details of the observations of CO($1-0$) line with ALMA, data reduction of them, and GMC identification are described in the paper. Thus, we give a brief summary here.

$^{12}$CO($1-0$) line (rest frequency: 115.271204~GHz) observations towards NGC~1300 were carried out as an ALMA Cycle 5 program with 12-m array (Program ID = 2017.1.00248.S, PI = F.Maeda). To cover the western bar, arm, and bar-end regions, two pointings were set. About 44 antennas were used with C43-5 configuration in which the projected baseline length ranged from 15.1~m to 2.5~km, corresponding to a maximum recoverable scale (MRS) of $\sim$21.4 arcsec at 115 GHz. Bandpass and phase were calibrated with J0423–0120 and J0340–2119, respectively.

Raw visibility data was calibrated by using the Common Astronomy Software Applications (\textsc{casa}) ver. 5.1.1. and the observatory-provided calibration script. We reconstructed the two-field mosaic image using \textsc{casa} ver. 5.4.0 using the \verb|multiscale| CLEAN algorithm \citep{Cornwell2008}
with Briggs weighting with robust = 0.5. 
We chose a velocity resolution of $5.0~\rm km~s^{-1}$. The resultant rms noise is 0.51 $\rm mJy~beam^{-1}$. We applied the primary beam correction on the output restored image and extracted the region within the primary beam correction factor smaller than 2.0. The data cube has an angular resolution of $0.44~{\rm arcsec} \times 0.30~{\rm arcsec}$, corresponding to $44~{\rm pc} \times 30~{\rm pc}$, and a pixel size of 0.12 arcsec.
Note that \citet{maeda_properties_2020} chose the above spatial and velocity resolution, pixel size, and line sensitivity to match the quality of the CO($1-0$) cube by \citet{colombo_pdbi_2014}, which studied the properties of GMCs in the prototype spiral galaxy of M51.

Using 3D clumps finding algorithm \textsc{cprops} \citep{rosolowsky_biasfree_2006}, which is designed to identify GMCs well even at low sensitivities, \citet{maeda_properties_2020} identified GMCs in NGC~1300 with almost the same parameters adopted in \citet{colombo_pdbi_2014}: \verb|threshold| = 4.0, \verb|edge| = 1.5, \verb|sigdiscont| = 0, \verb|bootstrap| = 50.
\textsc{cprops} identified 233 GMCs with the peak-to-noise ratio (S/N) above 4.
Fig.~\ref{fig:GMC_positions} shows the GMC distribution in the sky plane.
We detected 34, 119, and 49 GMCs in {\it bar}, {\it arm}, and {\it bar-end}, respectively. Here, definition of  environmental mask is represented as colour solid lines in Fig.~\ref{fig:GMC_positions}. \textsc{cprops} determines the size, velocity width, and flux of GMCs using moment methods. \textsc{cprops} corrects for the sensitivity by extrapolating GMC properties to those we would expect to measure with perfect sensitivity (i.e. 0 K) and the resolution by deconvolution for the beam and channel width.
The corrected mass, radius, and velocity dispersion are listed in the catalogue.
For a small GMC whose extrapolated size is smaller than the beam size, cataloged radius is 15.4 pc which corresponds to the beam size. The mass completeness limit is estimated to be $2.0 \times 10^5~\rm M_\odot$.
The median velocity dispersion of all GMCs is 5.2~$\rm km~s^{-1}$ and the median radius of resolved GMCs is 48.0~pc. There are no significant variations in the distribution of velocity dispersion and radius among the environments.
The marginally significant difference is seen in the distribution of molecular gas mass; the mass in {\it bar-end} is higher ($(5.0 - 15.5) \times 10^5~\rm M_\odot$) than those in {\it bar} ($(3.7 - 8.2) \times 10^5~\rm M_\odot$) and {\it arm} ($(3.4 - 9.7) \times 10^5~\rm M_\odot$).

\section{Estimated CCC speed}\label{sec: Estimated CCC speed}

In this section, we attempt to estimate the CCC speeds.
GMCs are considered to be moving along the global gas velocity field. Therefore, in Section~\ref{sec: Velocity Field Model}, we first obtain the global velocity field model of the molecular gases based on the CO($2-1$) 1st-moment map of NGC~1300 using an open tool, \textsc{diskfit} \citep{spekkens_modeling_2007, sellwood_quantifying_2010}. Then, in Section~\ref{sec: Estimation of the CCC speed}, we estimate GMC motion and the CCC speeds using the model and the catalogue. Section~\ref{sec:CCC Results}  shows the estimated CCC speed and the differences among the environments.
In Section~\ref{sec: Comparison of CCC speed with the simulation}, we compare the observations with the simulation by \citetalias{fujimoto_fast_2020}.

\subsection{Velocity Field Model}\label{sec: Velocity Field Model}
\subsubsection{1st-moment map of CO($2-1$) emission} \label{sec: CO(2-1)}
We made a $^{12}$CO($2-1$) (rest frequency: 230.538000~GHz) cube using the archival data which was observed with 12-m array and ACA (7-m + Total Power) under project 2018.1.01651.S (PI = A. Leroy) and 2015.1.00925.S (PI = B. Guillermo), respectively.
We calibrated raw visibility data using \textsc{casa} and the observatory-provided calibration script.
We imaged the interferometric cube using the CLEAN algorithm in \textsc{casa}.
We adopted a taper in the $uv$-plane to set the angular resolution of 4.0 arcesc, corresponding to 400 pc. 
Finally, using the \textsc{casa} task \verb|feather|, the interferometric cube
was combined with the cube obtained with the Total Power.
The rms noise of the data cube is $2.73~\rm mJy~beam^{-1}$ per 5.0 $\rm km~s^{-1}$ bin.

Fig.~\ref{fig:Diskfit_obs_model_res_fits}(a) shows the 1st-moment map of the CO($2-1$) in NGC~1300.
To make the image, we identified significant CO emissions using \textsc{cprops} as follows; we first identified 3D (position-position-velocity) regions with a signal-to-noise ratio (S/N) $\geq$ 4 in at least two adjacent velocity channels. Then, we expanded this mask to include all adjacent pixels with S/N $\geq 1.5$ in the 3D space. Using the masked cube, we made the 1-st moment map by the \verb|immoments| task in \textsc{casa}.

\subsubsection{Fitting with \textsc{diskfit}} \label{sec: Fitting with diskfit}

\begin{table}
 \caption{Fitted velocity components for NGC~1300.}
 \label{tab:Best fitted velocity}
 \begin{tabular}{ccc}
 \hline
 $r$ & $\bar{V}_t$ & $\bar{V}_{r}$  \\
 (kpc) & ($\rm km~s^{-1}$) & ($\rm km~s^{-1}$)   \\
  \hline
 0.4 & $-119.5 \pm 2.9$ & $  0.0 \pm  3.1$ \\
 1.2 & $-127.1 \pm 2.9$ & $  5.3 \pm  3.9$ \\
 2.0 & $ -86.5 \pm 4.0$ & $-18.3 \pm  6.1$ \\
 2.8 & $ -51.2 \pm 9.9$ & $ -0.3 \pm 18.2$ \\
 3.6 & $ -58.8 \pm 3.7$ & $ 43.7 \pm 13.2$ \\
 4.4 & $ -48.9 \pm 2.8$ & $ 30.3 \pm  6.4$ \\
 5.2 & $ -77.0 \pm 2.1$ & $ 20.1 \pm  4.4$ \\
 6.0 & $ -95.0 \pm 1.9$ & $ 18.3 \pm  2.5$ \\
 6.8 & $-113.5 \pm 1.8$ & $ 29.2 \pm  2.2$ \\
 7.6 & $-136.4 \pm 1.5$ & $ 27.4 \pm  1.7$ \\
 8.4 & $-150.0 \pm 1.6$ & $ 11.8 \pm  1.9$ \\
 9.2 & $-154.6 \pm 1.5$ & $ 14.4 \pm  1.7$ \\
10.0 & $-164.0 \pm 1.7$ & $  9.8 \pm  1.7$ \\
10.8 & $-165.9 \pm 1.7$ & $  6.3 \pm  2.5$ \\
11.6 & $-157.6 \pm 1.8$ & $ 19.7 \pm  2.8$ \\
12.4 & $-140.1 \pm 2.7$ & $ 39.2 \pm  4.0$ \\
13.2 & $-164.9 \pm 8.9$ & $ 34.5 \pm 10.0$ \\
 \hline
 \end{tabular}
\end{table}

\begin{figure}
	\includegraphics[width=\hsize]{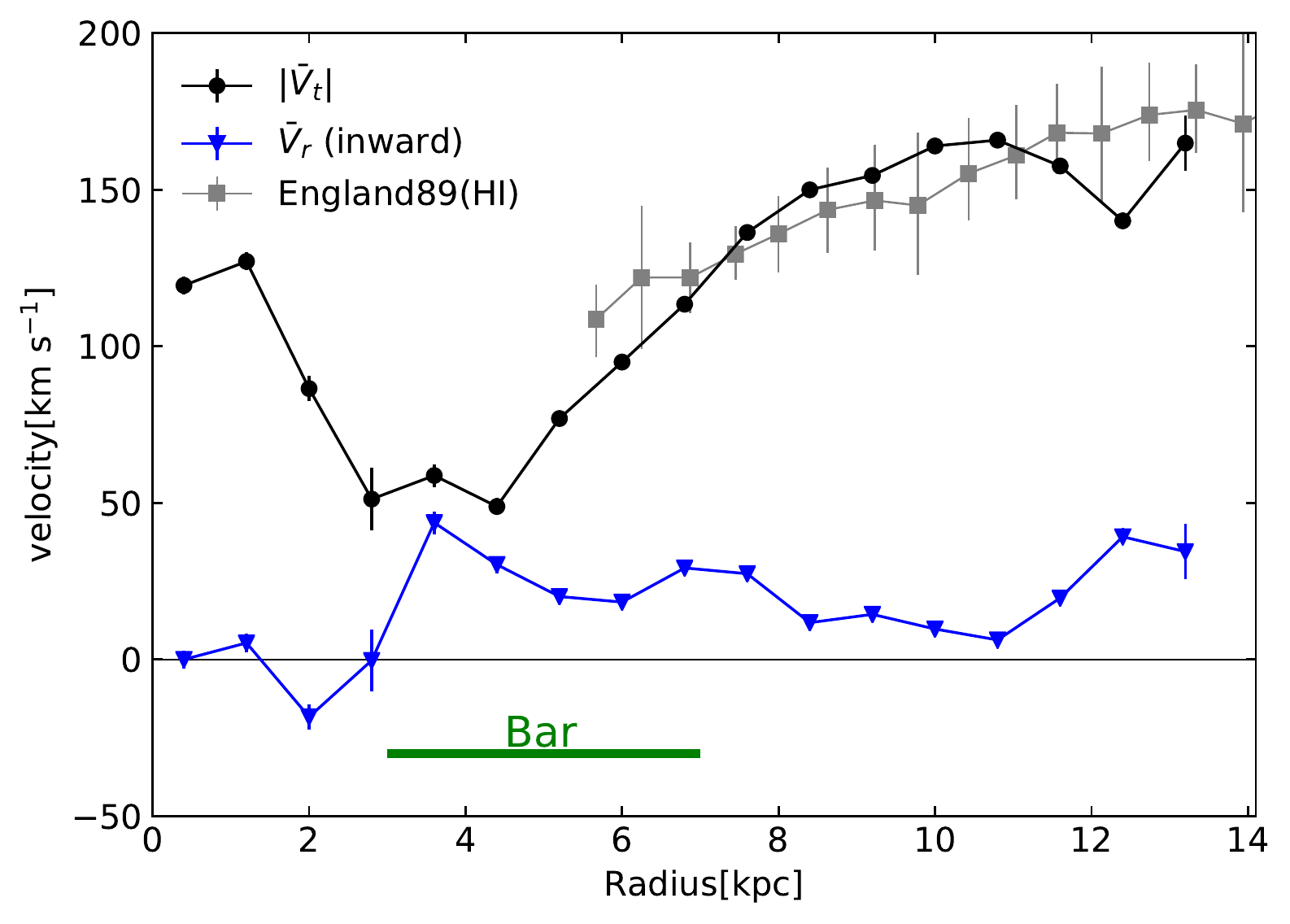}
    \caption{Fitted velocity components for NGC~1300. $\bar{V}_t(r)$ and $\bar{V}_r(r)$ are shown by the black circles and blue triangles, respectively. Gray squares show the angle-averaged rotation curve obtained from H\textsc{i} observations  by \citet{england_high-resolution_1989}. }
 \label{fig:rotation_curve_England}
\end{figure}

To determine the velocity field of galaxies, a number of techniques already exist.
Tilted-ring technique, which is introduced by \citet{Rogstad_1974ApJ...193..309R}
to describe the H\textsc{i} gas disc of M83, is now widely used.
This technique determines position angle, $\phi_d^\prime$, and inclination, $i$, as a function of galactocentric radius, and works well for smooth extended H\textsc{i} gas with the warp and radial inflow in the outer region of the galaxies \citep[e.g.][]{schmidt_radial_2016}. However, molecular gas is more compact and organized in a centrally thin disc, so the CO emissions are unlikely to trace such a wrap disc. Further, it is pointed out that $\phi_d^\prime$ and $i$ can be biased by applying the tilted-ring technique to kinematic tracer such as the CO emission in which non-circular motions are seen \citep[e.g.][]{lang_phangs_2020}. Therefore,  $\phi_d^\prime$ and $i$  are held constant with radius in this study. 

Assuming that molecular gases move in a 2D plane, the line-of-sight velocity in the sky plane, $V_{\rm obs}(\alpha, \delta)$, is the sum of the projected tangential, $V_t$, and radial, $V_r$, velocities in the disc plane:
\begin{equation}
    V_{\rm obs}(\alpha, \delta) = V_{\rm sys} + \sin i [ V_t(r, \theta) \cos \theta + V_r(r, \theta) \sin \theta ], 
\end{equation}
where $(r, \theta)$ is polar coordinate in the disc plane, $V_{\rm sys}$ is the systemic velocity, $i$ is the inclination.
This equation can be expressed as a Fourier series by
\begin{eqnarray}
   V_{\rm obs}(\alpha, \delta) = \nonumber \\
   V_{\rm sys} + \sin i [ \bar{V}_t(r) \cos \theta + \sum_{m = 1}^{\infty} V_{m, t} \cos \theta \cos(m \theta + \theta_{m, t}) \nonumber \\
   + \bar{V}_r(r) \sin \theta + \sum_{m = 1}^{\infty} V_{m, r} \sin \theta \cos(m \theta + \theta_{m, r}) ],
\end{eqnarray}
where the coefficients $V_{m, t}$ and $V_{m, r}$, and phases $\theta_{m, t}$ and $\theta_{m, r}$ are functions of $r$. The terms of $\bar{V}_t(r)$ and $\bar{V}_r(r)$ are the mean tangential and radial motion of the gas, respectively.
To estimate the global gas velocity, we used a purely axisymmetric kinematic model with coefficients of $m>0$ terms set to be zero, but retaining the $\bar{V}_r(r)$ term:
\begin{equation}
    V_{\rm obs}(\alpha, \delta) = V_{\rm sys} + \sin i [\bar{V}_t(r) \cos \theta + \bar{V}_r(r) \sin \theta].
    \label{eq: axisymmetric model}
\end{equation}

We performed fit to the line-of-sight velocity field (1st moment map) of CO($2-1$) with the Equation (\ref{eq: axisymmetric model}) using \textsc{diskfit}\footnote{\url{https://www.physics.queensu.ca/Astro/people/Kristine_Spekkens/diskfit/}} \citep{spekkens_modeling_2007, sellwood_quantifying_2010}.
In addition to the axisymmetric kinematic model expressed in Equation (\ref{eq: axisymmetric model}), non-axisymmetric kinematic models are implemented in \textsc{diskfit}. A bisymmetric model expresses the non-circular motion driven by barlike distortion using the $m = 2$ term in the Fourier expansion of $V_t$ and $V_r$. However, it is known that application of this model fails when the galactic bar is almost parallel to the major axis of the disc due to the degeneracy of the velocity components \citep{sellwood_quantifying_2010, randriamampandry_estimating_2015, randriamampandry_simulating_2018}.
Since the PA of the bar in NGC~1300 is close to that of the line of node (see Table~\ref{tab:NGC1300}), it is 
inappropriate to use the bisymmetric mode. Thus, we use the axisymmetric model of Equation (\ref{eq: axisymmetric model}).

\begin{figure}
\begin{center}
	\includegraphics[width=80mm]{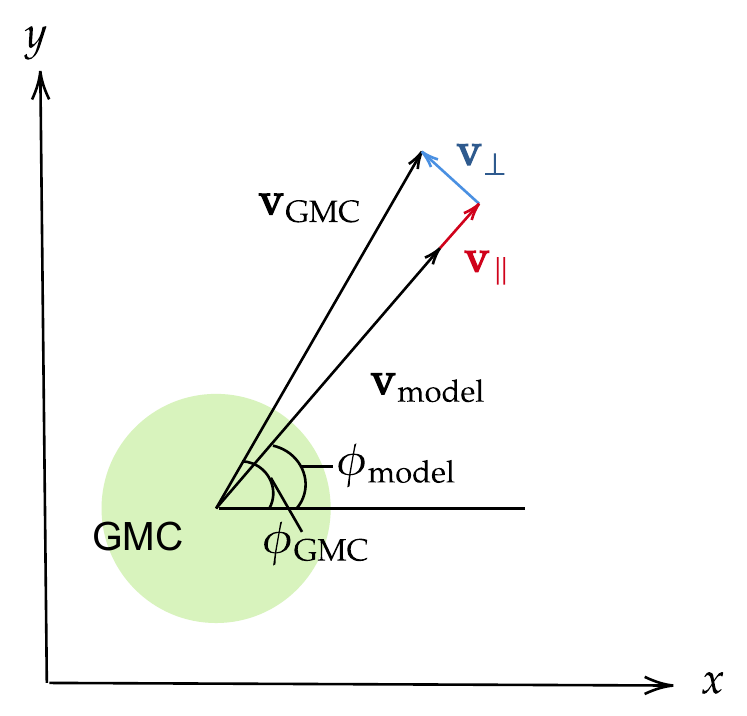}
    \caption{Schematic illustration showing the relationship between the velocity vector of a GMC, $\bm{V}_{\rm GMC}$, and the mean velocity field in the disc plane. 
    The $\bm{V}_{\rm model}$ is the velocity field obtained by the method
described in the text at that position. The $\phi_{\rm GMC}$ and $\phi_{\rm model}$ are angles of the projected major axis to $\bm{V}_{\rm GMC}$ and $\bm{V}_{\rm model}$, respectively. Difference vector between $\bm{V}_{\rm GMC}$ and $\bm{V}_{\rm model}$ is decomposed into the sum of two perpendicular vector components with one parallel to $\bm{V}_{\rm model}$, $\bm{V}_\parallel$, and one perpendicular to $\bm{V}_{\rm model}$, $\bm{V}_\perp$.
    }
 \label{fig:GMC_velocity}
 \end{center}
\end{figure}

\textsc{diskfit} consists of a set of model values for $\bar{V}_t$ and $\bar{V}_r$ around circular rings in the disc plane that project to ellipses on the sky plane with a common center, $(x_{\rm c}, y_{\rm c})$. Then, \textsc{diskfit} projects the model values and fit to the observed data, using linear interpolation to predict data values between rings. The radii of the rings set from 0.4 kpc to 13.2 kpc in a 0.8-kpc interval ($= 2 \times$ beam size) shown as gray dotted lines in Fig.~\ref{fig:Diskfit_obs_model_res_fits}. \textsc{diskfit} estimates the velocities $(\bar{V}_t, \bar{V}_r)$ and the other parameters ($V_{\rm sys}$, $\phi_d^\prime$, $i$, $x_{\rm c}$, $y_{\rm c}$) using a $\chi^2$ minimization technique. The values in Table~\ref{tab:NGC1300} were applied as their initial guesses. To estimate the uncertainty on the parameters, \textsc{diskfit} performs a bootstrap method and we made 100 times iterations.

Fig.~\ref{fig:Diskfit_obs_model_res_fits}(b) and (c) show the best-fitted model image and residual image, respectively.  
The best fitted disc parameters of $\phi_d^\prime$ and  $i$ are $281^\circ.8 \pm 0^\circ.5$ and $50^\circ.4 \pm 0^\circ.3$, respectively, and are almost same as the initial guesses derived by \citet{england_high-resolution_1989}.
There are a few observations that report different values from the result by \citet{england_high-resolution_1989} (e.g., $\phi_d^\prime = 267^\circ$ and  $i = 35^\circ$; \citealt{lindblad_velocity_1997}, $\phi_d^\prime = 278^\circ.1$ and  $i = 31^\circ.8$; \citealt{lang_phangs_2020}), but using these values as the initial guesses does not change the best-fitted values.
The best-fitted $V_{\rm sys}$ is $1544.7 \pm 0.2~\rm km~s^{-1}$. The best-fitted $\bar{V}_t$ and $\bar{V}_r$ are shown in Table~\ref{tab:Best fitted velocity}. Here, \textsc{diskfit} defines the positive direction of  $\bar{V}_t$ and $\bar{V}_r$ as counterclockwise and central direction, respectively. 

Fig.~\ref{fig:rotation_curve_England} shows the 
fitted tangential velocities (i.e. rotation curve) and radial velocities. In the bar region ($3~{\rm kpc} < r < 7~{\rm kpc}$), the radial velocity is large ($20 - 40~{\rm km~s^{-1}}$). This would be due to the elongated gas motion by the bar potential (\citetalias{fujimoto_fast_2020}). In the arm region ( $7~{\rm kpc} < r < 10~{\rm kpc}$), the gas motion is generally circular motion.
Our best fit rotation curve is consistent with that derived by  H\textsc{i} observations   \citep[][Fig.~\ref{fig:rotation_curve_England}]{england_high-resolution_1989}. 

\begin{figure}
\begin{center}
	\includegraphics[width=\hsize]{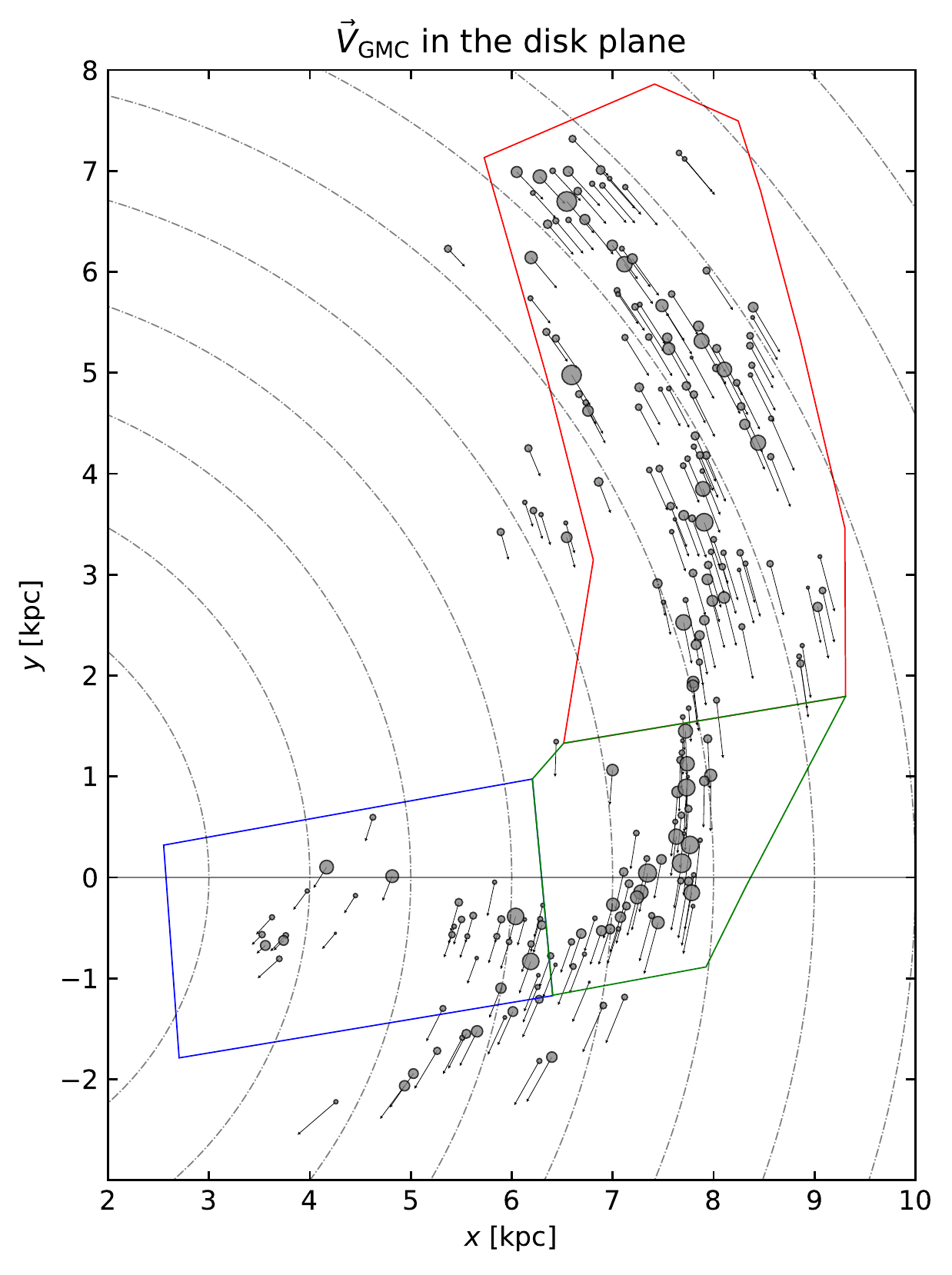}
    \caption{Derived $\bm{V}_{\rm GMC}$ in the disc plane based on the observed line-of-sight velocity of the GMCs and the modeled velocity field. The colour solid lines indicate the definition of the environmental mask (see Fig.~\ref{fig:GMC_positions}). Horizontal line shows the projected major axis. Dash-dotted lines shows concentric circles around the galactic center with 1~kpc interval.}
 \label{fig:GMC_vector_NGC1300}
 \end{center}
\end{figure}

\begin{table*}
 \caption{Estimated CCC speed in the different environments of NGC~1300}
 \label{tab:CCC speed}
 \begin{tabular}{lccccc}
  \hline
Region & $n_{\rm cloud}$ $^a$ & $n_{\rm ccc}$ $^b$ & $v_{\rm col}$ $^c$  & $n_{> 20 \rm km/s}$ $^d$& $f_{\rm ccc}$ $^e$\\
 & & & ($\rm km~s^{-1}$)  & ($\rm Gyr^{-1}~kpc^{-2}$) & ($\rm Gyr^{-1}$) \\
  \hline
 Bar     & 34 & 8  & $20.5^{+4.8}_{-5.7}$  & $51.6 \pm 6.5$ (50.0) & $24.8 \pm 2.5$ (23.5) \\
 Arm     & 119 & 12 & $11.2^{+3.5}_{-3.6}$ & $14.4 \pm 2.8$ (12.9)  & $10.7 \pm 0.5$ (10.2)  \\
 Bar-end & 49 & 12  & $16.0^{+11.3}_{-5.9}$ & $88.4 \pm 0.7$ (88.5) & $24.3 \pm 0.6$ (24.5) \\
 \hline
 \multicolumn{6}{l}{{\small $^a$ Number of GMCs.}} \\ 
 \multicolumn{6}{l}{{\small $^b$ Number of collisions.}} \\ 
 \multicolumn{6}{l}{{\small  $^c$ Estimated CCC speed based on the best-fitted model, which is noted as $M_{D25}^{D75}$, where $M$, }} \\ 
 \multicolumn{6}{l}{{\small  $D25$, and $D75$ are the median, the distance to the 25th percentile from the median, and }} \\
 \multicolumn{6}{l}{{\small  the distance  to the 75th percentile from the median of
 the number distribution, respectively.}} \\
 \multicolumn{6}{l}{{\small  $^d$ Number density of high speed CCC with $v_{\rm col} > 20~\rm km~s^{-1}$ per unit time
 and area. The  }} \\
 \multicolumn{6}{l}{{\small  uncertainty is calculated based on the 100 times iterations. The value in parentheses is derived  }} \\
  \multicolumn{6}{l}{{\small  using best-fitted model. }} \\
  \multicolumn{6}{l}{{\small  $^e$ Collision frequency. Notation is the same as $n_{> 20 \rm km/s}$.}}
 \end{tabular}
\end{table*}

\subsection{Estimation of the cloud-cloud collision speed}\label{sec: Estimation of the CCC speed}

To estimate the CCC speed, we first estimate the motion of the GMCs in the disc plane
using the mean velocity field of the molecular gas obtained in the previous subsection. Fig.~\ref{fig:GMC_velocity} is a schematic illustration showing the relationship between the velocity vector of a GMC, $\bm{V}_{\rm GMC}$, and the mean velocity field.
The $\bm{V}_{\rm model}$ is the velocity field obtained using \textsc{diskfit} at that position. The $\phi_{\rm GMC}$ and $\phi_{\rm model}$ are angles of the projected major axis (i.e. x axis) to $\bm{V}_{\rm GMC}$ and $\bm{V}_{\rm model}$, respectively. We decompose the  difference vector between $\bm{V}_{\rm GMC}$ and $\bm{V}_{\rm model}$ into the sum of two perpendicular vector components with one parallel to $\bm{V}_{\rm model}$, $\bm{V}_\parallel$, and one perpendicular to $\bm{V}_{\rm model}$, $\bm{V}_\perp$;
\begin{equation}
  \bm{V}_{\rm GMC} = \bm{V}_{\rm model} + \bm{V}_\parallel + \bm{V}_\perp.
\end{equation}
The observed line-of-sight velocity of the GMC, $V_{\rm GMC}^{\rm obs}$, is expressed as follows;

\begin{equation}
    V_{\rm GMC}^{\rm obs} = V_{\rm sys} + \sin i [|\bm{V}_{\rm model}| \sin \phi_{\rm model} + |\bm{V}_\parallel| \sin \phi_{\rm model} + |\bm{V}_\perp| \cos \phi_{\rm model}].
\end{equation}
Using the GMC catalogue by \citet{maeda_properties_2020} and the modeled mean velocity field, we calculated $|\bm{V}_\parallel|$ and then, derived $ \bm{V}_{\rm GMC}$. Here, we assume the direction of the GMC motion is parallel to the global velocity field i.e. $\bm{V}_\perp = \bm{0}$ ($\bm{V}_{\rm GMC} \parallel \bm{V}_{\rm model}$. See Section~\ref{Biases on GMC motion} for more details.).

Fig.~\ref{fig:GMC_vector_NGC1300} shows the distribution of $\bm{V}_{\rm GMC}$ in the disc plane of NGC~1300. The galactic center is set to be $(x_{\rm c}, y_{\rm c}) = (0,0)$.
The motion of GMCs in {\it bar} deviates from the circular motion. This would be due to the elongated elliptical gas motion shifted by the bar potential.
Such deviation can be seen in {\it bar-end}.
On the other hand, GMCs in {\it arm} are close to the circular orbit around the galactic centre. Similar motions are seen in the simulation modeled NGC~1300 ( \citetalias{fujimoto_fast_2020}; see Fig. 10).

Next step is to estimate the CCC speed.
We denote the velocity vectors of two GMCs located at $(x_i, y_i)$ and $(x_j, y_j)$ at the current time $(t = 0)$ as $\bm{V}_{\rm GMC}^i = (v_{x,i},v_{y,i})$ and $\bm{V}_{\rm GMC}^j = (v_{x,j},v_{y,j})$, respectively. To estimate the collision velocity, we impose some assumptions:
(1) All GMCs are spherical shape. (2) GMCs are in constant linear motion. (3) The interactions among clouds such as gravity and viscosity are negligible. Thus, the distance between the two GMCs at time of $t$, $D_{i,j}(t) $, becomes
\begin{eqnarray}
  D_{i,j}(t) = \nonumber \\
\left\{ [(x_i + v_{x,i} t) - (x_j + v_{x,j} t)]^2 
+[(y_i + v_{y,i} t) - (y_j + v_{y,j} t)]^2 \right\}^{1/2}. 
\end{eqnarray}
To identify the potentially colliding pairs, we first extracted the pairs $(i,j)$ which are within a distance of 300 pc and  move towards each other;
$D_{i,j}(0) \leq 300~{\rm pc}$ and $\frac{dD_{i,j}}{dt}|_{t =0} < 0$. Here, we set a cut-off distance of 300~pc to include 75 percent of the distances to the nearest GMC in {\it bar}, where the GMCs are most sparsely distributed (see also Fig.~\ref{fig:velocity_deviation_in_NGC1300}(a)). Next, we extract the pairs such that there exists a $t$ that satisfies $D_{i,j}(t) \leq R_i+R_j$ in the range of $0 \leq t < 10~{\rm Myr}$, where $R_i$ and $R_j$ are catalogued radii.
If there are other GMCs between the two GMCs (ID = $i,j$), we exclude the pair. Finally, we determine the relative speed as the estimated CCC speed; 
\begin{equation}
    v_{\rm col} = \left|\bm{V}_{\rm GMC}^i - \bm{V}_{\rm GMC}^j \right|.
\end{equation}
It should be noted that the assumptions of (2) and (3) lead to neglect of the impact of collisionless encounters (scattering by other GMCs) in this study. Therefore, there is a possibility that the paths of the pairs expected to collide with each other in the near future may change by the collisionless encounters.

We limited the range of $t$ within 10~Myr because the assumption of straight motion for GMCs is a reasonable approximation of the circular motion  within this range. In other words, the moving distance is small enough for around the galaxy; the $\omega t$ is less than 0.2 rad, which corresponds to the fraction of the orbital time of $\sim$1/30, within 10~Myr. Here, $\omega$ is the angular velocity of $\bar{V}_t /r \sim 0.02~\rm rad~ Myr^{-1}$, where $\bar{V}_t$ and $r$ in {\it arm} are $\sim \rm 150~km~s^{-1}$ and $r \sim 9$~kpc. Note that the timescale of 10~Myr ($\sim$1/30 of the orbital time) is comparable or shorter than the collision timescales obtained from observations and galaxy simulations; $\sim 7~\rm Myr$ in the W43 GMC complex in the Milky Way's bar-end region \citep{kohno_forest_2020}, $\sim25~\rm Myr$ ($\sim 1/5$; \citealt{tasker_star_2009}),  $8-10~\rm Myr$ ($\sim 1/15$) in a simulation with spiral arms and $\rm 28~Myr$ ($\sim 1/5$) with no imposed spiral arms \citep{dobbs_frequency_2015}, and  71~Myr ($\sim 1/3$), 57~Myr ($\sim 1/5$), and 72~Myr ($\sim 1/4$) in {\it bar}, {\it bar-end}, and {\it arm}, respectively, in a simulation of a strongly barred galaxy (\citetalias{fujimoto_fast_2020}).



\subsection{Results}\label{sec:CCC Results}
\begin{figure}
\begin{center}

	\includegraphics[width=\hsize]{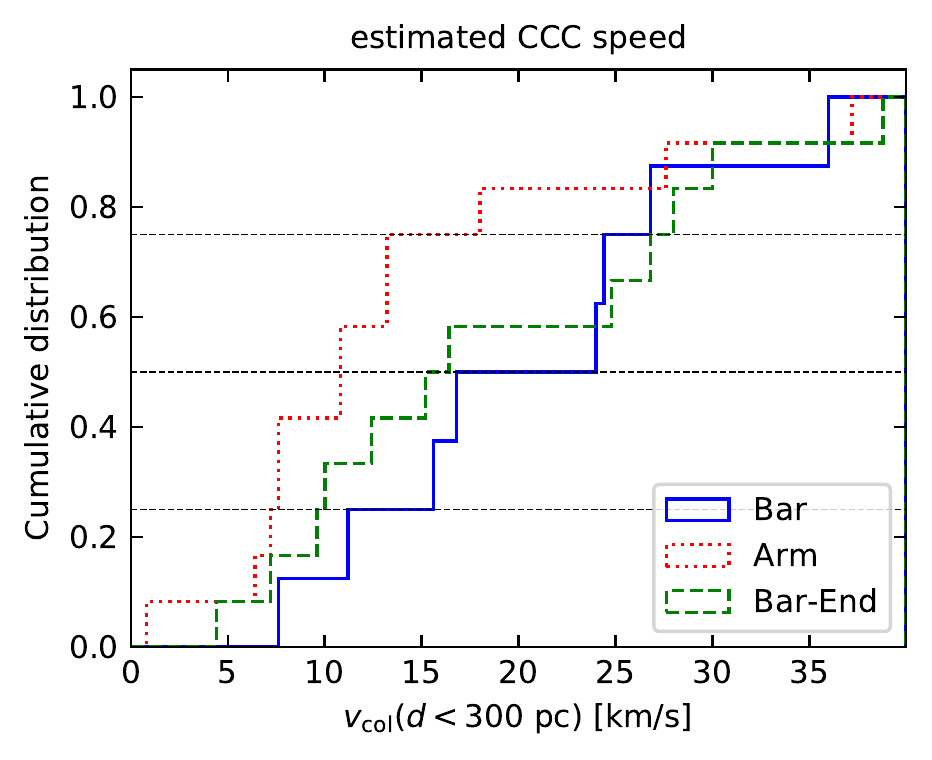}
    \caption{Normalized cumulative distribution function of the CCC speed. The samples are the potentially colliding pairs we identified (see text). The horizontal lines show the cumulative distribution function of 0.25, 0.50, and 0.75, respectively.}
 \label{fig:collision_velocity_in_NGC1300}
 \end{center}
\end{figure}

We find a tendency for the estimated CCC speed, $v_{\rm col}$, to depend on the environment. Fig.~\ref{fig:collision_velocity_in_NGC1300} shows the normalized cumulative distribution function of the estimated CCC speed. The GMCs in the samples are expected to collide each other within 10~Myr from the current time. As shown in Table~\ref{tab:CCC speed}, the $v_{\rm col}$ in {\it bar} and {\it bar-end} are higher than that in {\it arm}: $14.8 - 25.3~\rm km~s^{-1}$ in {\it bar}, $10.1 - 25.3~\rm km~s^{-1}$ in {\it bar-end}, and $7.6 - 14.7~\rm km~s^{-1}$ in {\it arm}, respectively.
This result does not change by using the fitting results of the 100 times bootstrap iterations (Section~\ref{sec: Fitting with diskfit}); 
median value of the $v_{\rm col}$ is $19.9 \pm 2.3~\rm km~s^{-1}$, $16.1 \pm 0.2~\rm km~s^{-1}$, and $11.1 \pm 0.2~\rm km~s^{-1}$ in {\it bar}, {\it bar-end}, and {\it arm}, respectively.

We statistically checked the environmental variation of the $v_{\rm col}$ distribution using the two-sided Kolmogorov-Smirnov (K-S) test.
We used \verb|stats.ks_2samp| function of python's Scipy package, which calculates $p$-value based on the approximately formula given by \citet{Stephens:1970ic}.
Following the conventional criteria, the two cumulative distribution functions are considered to be significantly different if the $p$-value is less than 0.01. For a $p$-value within 0.01 to 0.05, the difference between the two distribution is considered to be marginally significant. The $p$-value showing significant difference was not detected; $>0.1$. Although this result indicates that there would be no statistically significant difference among the environments, it is possible that the $p$-value may be high due to the small number of samples. Generally, when comparing two samples (the number of samples of $N_1$ and $N_2$), the effective number of sample $N_{\rm e} = N_1 N_2 / (N_1 + N_2)$ is preferably more than 8 in order to obtain the accurate $p$-value according to the approximation by \citet{Stephens:1970ic}. However, in this study, $N_{\rm e}$ is 4.8, 4.8, and 6.0 when comparing {\it bar} and {\it arm}, {\it bar} and {\it bar-end}, and {\it arm} and {\it bar-end}, respectively. Therefore, it is premature to draw a statistical  conclusion about the environmental dependence on $v_{\rm col}$ and further observations are desirable (e.g. observations of the GMC across the whole NGC~1300.).

\begin{figure}
\begin{center}
	\includegraphics[width=\hsize]{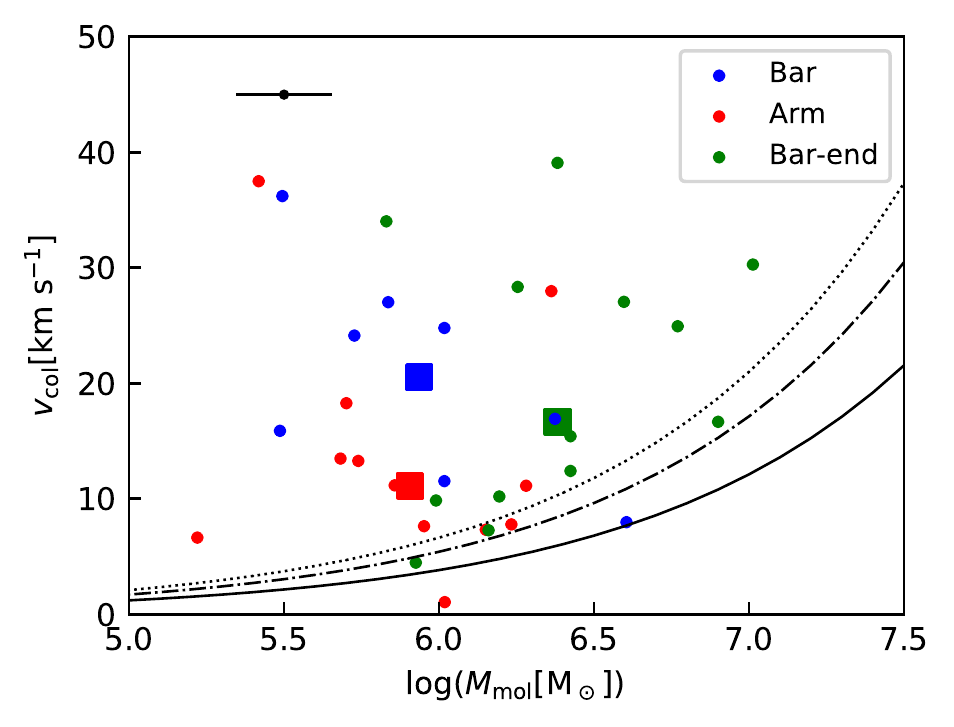}
    \caption{Scatter plot of the CCC speed versus the mass of the most massive GMC in each collision. The $v_{\rm col}$  was derived  using the best-fitted velocity-field model. The large symbols show the median values. The black bar in the upper left shows the typical uncertainty of the molecular gas mass of the GMCs derived by \citet{maeda_properties_2020}. The black lines show a free fall speed  as a function of GMC mass (see equation (\ref{eq:free-fall})). We set $R = 50~\rm pc$ and $D_{\rm nearest} = 120~\rm pc$ (solid line), $R = 50~\rm pc$ and $D_{\rm nearest} = 150~\rm pc$ (dash-dotted line), and $R = 50~\rm pc$ and $D_{\rm nearest} = 200~\rm pc$ (dotted line).
    }
 \label{fig:Mmol_vs_Vcol}
 \end{center}
\end{figure}

Fig. \ref{fig:Mmol_vs_Vcol} shows the scatter plot between the estimated CCC speed and the molecular gas mass of the most massive GMC in each pair. 
Large symbols show the median values in the environments.
Comparing {\it bar} and {\it arm}, molecular gas mass is comparable but the $v_{\rm col}$ in {\it bar} tends to be higher than that in {\it arm}. On the other hand, comparing {\it bar} and {\it bar-end}, $v_{\rm col}$ is roughly comparable but the molecular gas mass in {\it bar-end} tends to be higher than that in {\it bar}. In Sections \ref{sec: The cause for the different star formation activity between bar and bar-end with the same high CCC speed} and \ref{sec: The cause of the low SFE in the bar region}, we further discuss the connection among star formation activity, CCC speed, and molecular gas mass of the colliding GMCs.

Next, we focus on high-speed CCCs with $v_{\rm col} > 20~\rm km~s^{-1}$. Such a high-speed CCC can shorten the gas accretion phase of the cloud cores formed, leading to massive star formation suppression \citep{takahira_cloud-cloud_2014, takahira_formation_2018}. As shown in Table~\ref{tab:CCC speed}, we find that the number density of high speed CCC with $v_{\rm col} > 20~\rm km~s^{-1}$ per unit time and area, $n_{> 20 \rm km/s}$, is significantly higher in {\it bar} ($\sim 52~\rm Gyr^{-1}~kpc^{-2}$) and {\it bar-end} ($\sim 88~\rm Gyr^{-1}~kpc^{-2}$)  than in {\it arm} ($\sim 14~\rm Gyr^{-1}~kpc^{-2}$). High number density in both {\it bar} and {\it bar-end}, where the star formation activity is significantly different, suggests that whether or not star formation is induced by CCC depends not only on the CCC speed, but also on other parameters. \citet{takahira_formation_2018} pointed out that the core mass growth also depends on the mass (size) of the colliding clouds.
This point is discussed in more detail in Section~\ref{sec: The cause for the different star formation activity between bar and bar-end with the same high CCC speed}.

Lastly, we investigated collision frequency defined as,
\begin{equation}
    f_{\rm ccc} = \cfrac{n_{\rm ccc}}{n_{\rm cloud} \Delta t},
\end{equation}
where $n_{\rm ccc}$ is the number of collisions, $n_{\rm cloud}$ is the number of clouds in each environment (Table~\ref{tab:CCC speed}), and $\Delta t$ is the tracking time of 10~Myr. The collision frequency is estimated to be $10 -25~{\rm Gyr^{-1}}$ (Table~\ref{tab:CCC speed}). We find that the collision frequencies in {\it bar} and {\it bar-end} are about two times higher than that in {\it arm}. The lack of star formation activity despite the high collision frequency in {\it bar} suggests that the star formation activity is not controlled only by the collision frequency.

We checked the dependence on the cut-off distance. If the cut-off distance is varied from 200 pc to 400 pc, the number of samples in each environment increases or decreases by a few, and the results remain almost the same. In the case of 200~pc, $v_{\rm col}$ in {\it bar}, {\it bar-end}, and {\it arm} are $20.0^{+4.6}_{-7.4}~\rm km~s^{-1}$, $15.4^{+10.6}_{-5.4}~\rm km~s^{-1}$, and $11.2^{+3.5}_{-3.6}~\rm km~s^{-1}$, respectively.
In the case of 400~pc, $v_{\rm col}$ in {\it bar}, {\it bar-end}, and {\it arm} are $20.5^{+4.8}_{-5.7}~\rm km~s^{-1}$, $16.7^{+11.7}_{-6.5}~\rm km~s^{-1}$, and $11.2^{+3.5}_{-3.6}~\rm km~s^{-1}$, respectively.
However, if we take the cut-off distance as 150 pc or less, we can hardly see the environmental dependence in the CCC speed. This is because the number of pairs halved in the bar region. Since the median value of the distance to the nearest GMC is ~150 pc in the bar (see also Fig.\ref{fig:velocity_deviation_in_NGC1300}(a)), setting the cut-off distance to 150 pc or less would be inappropriate, leading to an unfair decrease in the number of the sample in the bar. 


\subsection{Comparison of estimated CCC speeds with the simulation by \citet{fujimoto_fast_2020}}\label{sec: Comparison of CCC speed with the simulation}

Here we compare the CCC speed between  our observations and  simulation by \citetalias{fujimoto_fast_2020}.
\citetalias{fujimoto_fast_2020} presented a hydrodynamical simulation of a strongly barred galaxy.
The simulation was run using adaptive mesh refinement hydrodynamics code \textsc{enzo} \citep{bryan_enzo_2014}. The minimum cell size is $\sim2$~pc. They used a fixed potential model which is taken from observational results of NGC~1300 \citep{england_dynamical_1989}. There is no star formation or stellar feedback in the simulation.

For the CCC speed in {\it bar}, the simulation result is generally consistent with the estimated CCC speed. In the simulation, the fraction of collided clouds with CCC speed more than 20~$\rm km~s^{-1}$ is roughly 40 percent in {\it bar}, which is also consistent with our results. In {\it arm}, the simulation result is median value is $\sim 10~\rm km~s^{-1}$, which is comparable to our results. However, the results in {\it bar-end} are different; estimated CCC speed is higher than that in the simulation ($\sim 10~\rm km~s^{-1}$). This result indicates that the deviation from the circular motion may be larger in {\it bar-end} than that expected in the simulation.

In Fig. \ref{fig:Mmol_vs_Vcol}, the black lines show the free-fall speed as a function of the GMC mass. The free-fall speed is defined in equation (13) in \citetalias{fujimoto_fast_2020} as,
\begin{equation}
\label{eq:free-fall}
    v_{\rm ff} = \sqrt{2 G M_{\rm mol} \left( \frac{1}{2R} - \frac{1}{D_{\rm nearest}} \right)},
\end{equation}
where $M_{\rm mol}$ is the molecular gas mass of the GMC, $R$ is the radius of the GMC, and $D_{\rm nearest}$ is the distance to the nearest GMC. We use  $R = 50~\rm pc$, which is the median value in the catalogue by \citet{maeda_properties_2020}, and  $D_{\rm nearest} = $ 120, 150, and 200~pc. In the simulation by \citetalias{fujimoto_fast_2020}, the {\it bar-end} and {\it arm} clouds roughly follow the free-fall line, which supports the idea of a large contribution of gravitational interactions between clouds to driving CCCs. However, most of the observed values in all environments are located above the lines. This suggests that gravitational interactions have only a small contribution to driving collisions. As mentioned in the next section, not the gravitational interaction but the random-like motion of GMCs may contribute to the collisions to some extent.

For the collision frequency, unlike the observations, a significant environmental dependence is not seen in the simulation. The collision frequencies in the simulation in {\it bar} and {\it bar-end} ($\sim 15~\rm Gyr^{-1}$) are slightly lower than those in the observations. The possible reason for the difference is that the size of the clouds in the simulation is smaller than that of the observed GMCs. Because of differences in the threshold density for cloud identification and the spatial resolution (see Section~\ref{sec:Caveats on the comparison between observation and simulation}), the cloud radius is smaller in the simulation ( median of $\sim 8$ pc) than that of observed GMCs ($\sim 40$ pc). Thus, the cross-sections of clouds in the simulation may be smaller and then collision frequencies get lower.

\section{velocity deviation}\label{sec: velocity deviation}

We investigated the properties of CCC with another approach using  GMC catalogue as a model-independent method. We examined the distance to the nearest GMC, velocity deviation between the GMC and its surrounding GMCs, and collision frequency of GMCs by assuming that GMCs are in random motions.

\begin{figure}
	\includegraphics[width=\hsize]{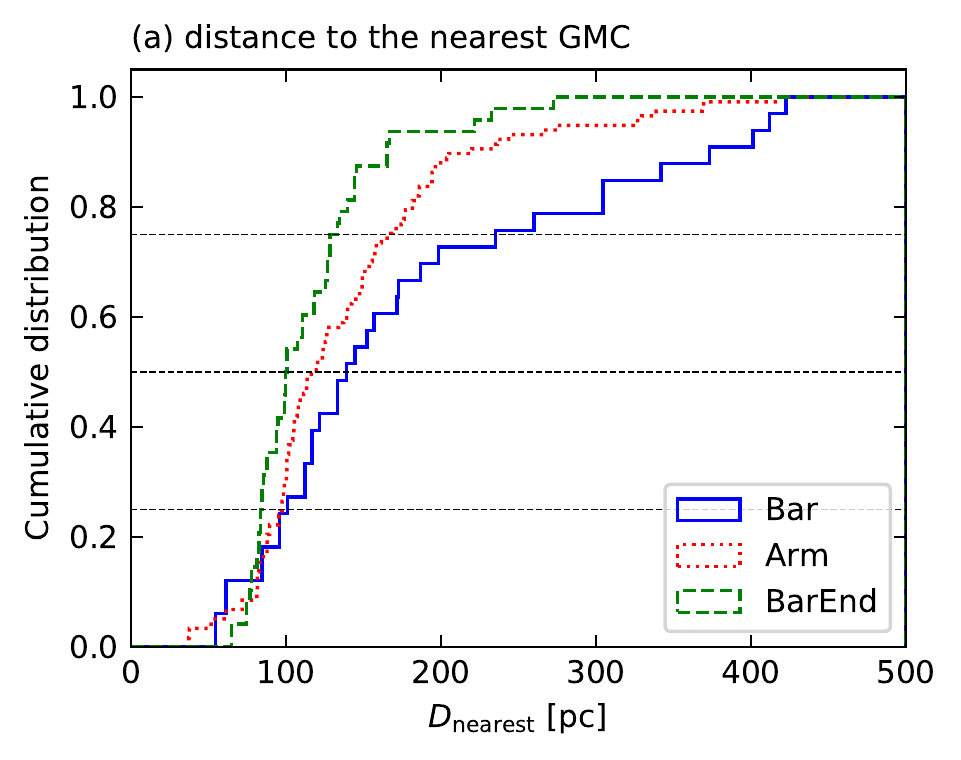}
	\includegraphics[width=\hsize]{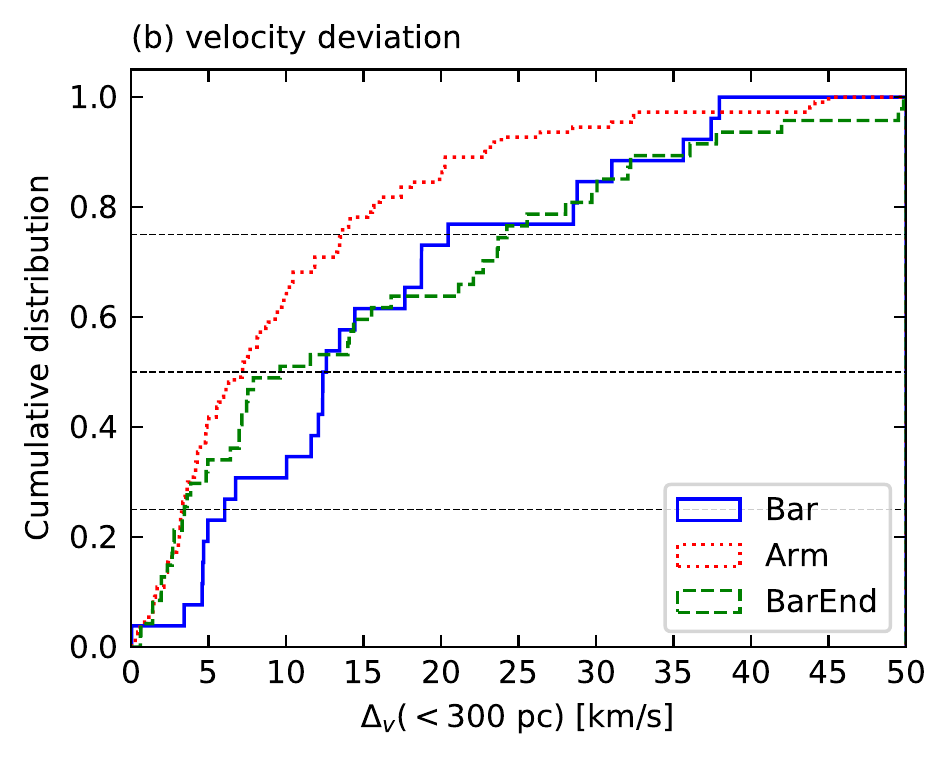}
	\includegraphics[width=\hsize]{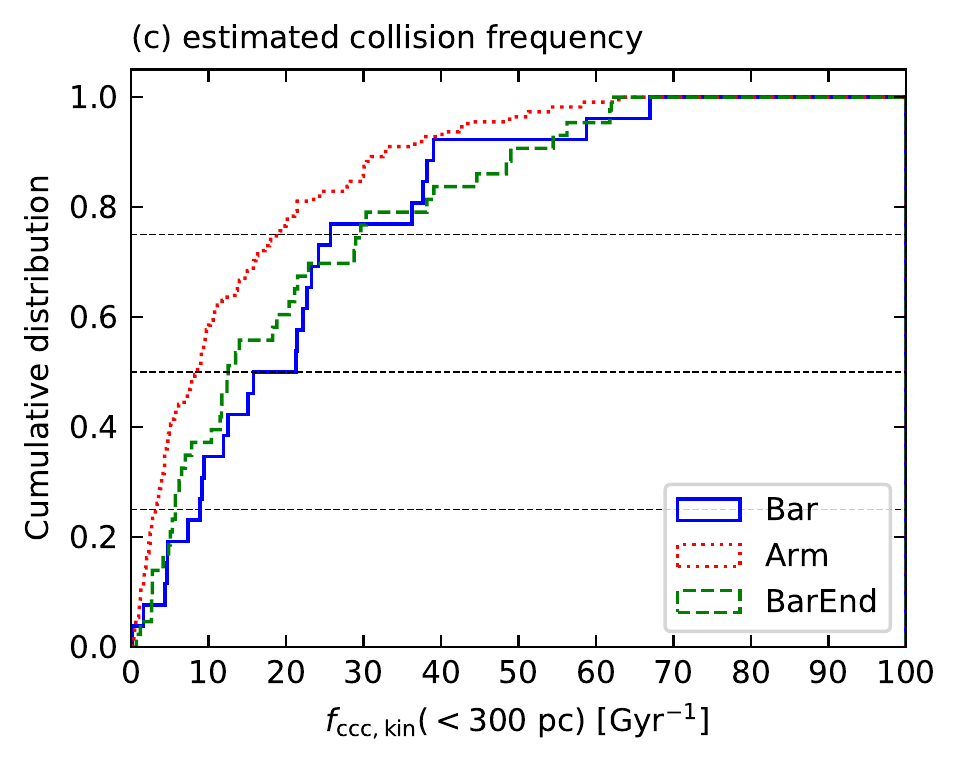}
    \caption{Normalized cumulative distribution function of (a) the distance to the nearest GMC, (b) the velocity deviation between the GMC and its surrounding GMCs within a radius of $r < 300~\rm pc$ and, (c) the estimated collision frequency of GMCs.}
 \label{fig:velocity_deviation_in_NGC1300}
\end{figure}

Fig.~\ref{fig:velocity_deviation_in_NGC1300}(a) shows the normalized cumulative distribution function of the distance to the nearest GMC.
The median value in {\it bar}, {\it arm}, and {\it bar-end} are 142, 120, and 100 pc, respectively; the {\it bar-end} GMCs have a smaller distance to the nearest GMC. This result indicates that the spatial distribution of the {\it bar-end} GMCs is more crowded, which would be one of the causes of the high collision frequency.
The difference between observations and the simulations by \citetalias{fujimoto_fast_2020} is only a few tens of parsec, but unlike observations, the {\it bar} clouds have a smaller distance to the nearest cloud than other regions in the simulation. As we explain in the later sections, the differences can be attributed to the differences in methodology of cloud identification and in spatial resolution.

Next, we calculated the velocity deviation between the GMC and its surrounding GMCs defined as, 
\begin{equation}
    \Delta_v (<r) = \sqrt{2} \left|v_{\rm los} - \bar{v}_{\rm los}(<r) \right|/ \sin i,
\end{equation}
where the $v_{\rm los}$ is the line-of-sight velocity of the GMC and $\bar{v}_{\rm los}(<r)$ is the mass-weighted mean line-of-sight velocity of its surrounding GMCs within a radius of $r$, respectively, and $i$ is the inclination of NGC~1300. We multiplied by the factor of $\sqrt{2}$ by assuming that the GMCs move randomly in a 2D plane. We set a cut-off distance of 300 pc to include 75 percent of the distances to the nearest GMC in the bar region (Fig.~\ref{fig:velocity_deviation_in_NGC1300}(a)). The simulation by \citetalias{fujimoto_fast_2020} shows that the velocity deviation in the bar region is larger than that in the arm and bar-end regions. This tendency is the same as the CCC speed in the simulation, indicating that the velocity deviation can be a good indicator of the qualitative trends of the CCC speed.

\begin{table}
 \caption{Velocity deviation ($r = 300$ pc) in the different environments of NGC~1300}
 \label{tab:Velocity deviation}
 \begin{tabular}{lccc}
  \hline
Region & \# $^a$ & $\Delta_v$ $^b$& $f_{\rm ccc, kin}$ $^b$\\
 &  & ($\rm km~s^{-1}$) & ($\rm Gyr^{-1}$) \\
  \hline
 Bar     & 26  & $12.5^{+7.5}_{-6.3}$  & $18.6^{+6.8}_{-9.6}$\\
 Arm     & 111 & $7.2^{+6.4}_{-3.9}$  & $8.4^{+10.3}_{-5.1}$\\
 Bar-end & 48  & $10.6^{+14.1}_{-7.0}$ & $16.1^{+24.3}_{-10.0}$ \\
 \hline
  \multicolumn{4}{l}{{\small  $^a$  Number of the GMC which has the }}\\
  \multicolumn{4}{l}{{\small  nearest GMC within 300~pc.}} \\
  \multicolumn{4}{l}{{\small  $^b$  Notation is the same as $v_{\rm col} $ in Table~\ref{tab:CCC speed}.}}
 \end{tabular}
\end{table}

Table~\ref{tab:Velocity deviation} shows the observed $\Delta_v (< 300~{\rm pc})$ in each environment. Fig.~\ref{fig:velocity_deviation_in_NGC1300}(b) shows the cumulative distribution of $\Delta_v (< 300~{\rm pc})$. We find that the $\Delta_v$ in {\it bar} and {\it bar-end} are larger than that in {\it arm}; $\Delta_v (< 300~{\rm pc})$ in {\it bar} and {\it bar-end} are $6.2 - 20.0~\rm km~s^{-1}$ and $3.6 - 24.7~\rm km~s^{-1}$, respectively, followed by $3.3 - 13.6~\rm km~s^{-1}$ in {\it arm}. This result suggests that the  CCC speeds in {\it bar} and {\it bar-end} are higher than that in {\it arm}, which is consistent with the collision speed derived in Section~\ref{sec:CCC Results}. We statistically checked the environmental variation of the $\Delta_v (< 300~{\rm pc})$ distribution using  the two-sided K-S test. Significant differences are seen between  {\it bar} and {\it arm}, and {\it bar-end} and {\it arm}; The $p$-values are  $0.011$ and $0.018$, respectively. On the other hand, there is no significant difference between {\it bar} and {\it bar-end}; $p$-value is 0.37.  
We checked the dependence on the cut-off distance. 
The tendency of environmental dependence does not change even if the cut-off distance is varied from 200~pc to 400~pc;
$\Delta_v(<200~{\rm pc})$ in {\it bar}, {\it bar-end}, and {\it arm} are $4.7 - 22.8~\rm km~s^{-1}$, $6.7 - 26.6~\rm km~s^{-1}$, and $1.0 - 13.4 ~\rm km~s^{-1}$, respectively.
$\Delta_v(<400~{\rm pc})$ in {\it bar}, {\it bar-end}, and {\it arm} are $7.3 - 18.6~\rm km~s^{-1}$, $6.3 - 25.9~\rm km~s^{-1}$, and $4.5-16.3~\rm km~s^{-1}$, respectively.
Although the trend of environmental dependence is similar, the estimated CCC speed is larger than the velocity deviation. As described in \citetalias{fujimoto_fast_2020}, this might be because we use only GMCs which are the potentially colliding pairs to estimate the CCC speed, while we use all GMCs to calculate the velocity deviation including those that have low-velocity deviation and are unlikely to collision soon.

Lastly, we estimated the collision frequency of the GMCs using cloud's kinematics model using the formula applied by \citetalias{fujimoto_fast_2020} defined as,
\begin{equation}
    f_{\rm ccc,kin} = 2 R N \Delta_v,
\end{equation}
where $R$ is the catalogued radius of the GMC, $N$ is the surface number density of GMCs. Here, five assumptions are imposed: (1) GMCs are spherical shape and which is much smaller than the average distance between them. (2) GMCs are in constant, rapid, and random motion. (3) GMCs undergo random elastic collisions between themselves.
(4) Except during collisions, the interactions among GMCs such as gravity and viscosity are negligible. (5) GMCs move in a 2D plane. This model is well known as kinetic theory and is often applied to ideal gases.
The $N$ and $\Delta_v$ are calculated within a circle whose radius is 300 pc centered at the GMC. 

Fig.~\ref{fig:velocity_deviation_in_NGC1300}(c) shows the normalized cumulative distribution function of the estimated collision frequency, $f_{\rm ccc,kin}$, of GMCs. The $f_{\rm ccc,kin}$ and $f_{\rm ccc}$ derived in Section~\ref{sec:CCC Results} are roughly comparable. It is natural to get this result because we made almost the same assumptions in calculating both frequencies. Compared to the observations and simulation by \citetalias{fujimoto_fast_2020}, $f_{\rm ccc,kin}$ in {\it bar} is comparable. \citetalias{fujimoto_fast_2020} suggests that a random-like motion of clouds induced by the elongated gas stream due to the bar potential could be the physical mechanism of the high-speed CCC in {\it bar}. \footnote{  \citet{fujimoto_environmental_2014} suggested another mechanism; massive clouds create strong gravitational interactions and thus increase the CCC speed. However, the mass of the massive clouds exceeds $10^7~\rm M_\odot$, which is inconsistent with the observed GMCs. That might be because they performed the GMC formation simulation for over 200~Myr without stellar feedback. In addition, their galaxy model is the intermediate-type barred galaxy M83, not the strongly barred galaxy NGC~1300 we discuss in this paper. That also might be the reason \citet{fujimoto_environmental_2014} did not see the strongly elongated gas stream. } 
In the simulation, the median $f_{\rm ccc,kin}$ in the {\it bar-end} and {\it arm} are $\sim 5~\rm Gyr^{-1}$, which is much lower than those in the observations. The reason of this difference is that the observed $\Delta_v$ is much  larger than that in the simulation ($\sim 2~\rm km~s^{-1}$). \citetalias{fujimoto_fast_2020} concludes that gravitational interactions between clouds have a larger contribution to collisions in {\it arm} and {\it bar-end} regions than random-like motion of GMCs, resulting in low $f_{\rm ccc,kin}$.
However, based on the observed $f_{\rm ccc,kin}$, random-like motion of GMCs may have some extent contribution to the collisions in {\it arm} and {\it bar-end}.

\section{Discussion}\label{sec:Discussion}

\subsection{Biases and caveats}\label{sec: Biases and caveats}

\subsubsection{Biases on GMC motion}\label{Biases on GMC motion}
In this study, several assumptions are imposed on the GMC motion, and we discuss them here. First, we neglect the vertical GMC distributions in the galactic mid-plane. This assumption would have little effect on the above results. In the simulation by \citetalias{fujimoto_fast_2020}, all clouds lie in the small region between $-30~{\rm pc} < z < 30~{\rm pc}$ around the galactic mid-plane. In the Milky Way, the median height of GMC from Galactic plane is $\sim 30$~pc \citep{solomon_mass_1987}. These results indicate we can neglect the effect of the vertical GMC distribution.

Second assumption is that we neglect the vertical velocities of GMC motion.
In the simulation by \citetalias{fujimoto_fast_2020}, the vertical velocity is much less than $\rm 10~km~s^{-1}$ and any difference among the environments is not seen. Therefore, the vertical motion of the GMCs can be neglected.

Third assumption is that $\bm{V}_{\rm GMC}$ is parallel to $\bm{V}_{\rm model}$, i.e. $\bm{V}_\perp = \bm{0}$. In other words, we neglected the impact of collisionless encounters (scattering by other GMCs), which increases the radial perturbation of the GMCs and thus increases the collision frequency. Although it is impossible to derive the $\bm{V}_\perp$ with current data, we derived the $v_{\rm col}$  assuming the $\bm{V}_\perp$ to check this assumption. We set the $|\bm{V}_\perp|$ of each GMC randomly selected from a Gaussian distribution with a mean of $0~\rm km~s^{-1}$ and a standard deviation of $7~\rm km~s^{-1}$, which is comparable to the cloud-cloud velocity dispersion within 3~kpc of the Sun \citep{stark_kinematics_1989}. We made 100 times iterations.
As a result, the collision frequency in the arm region increased by a factor of 1.5 because the deviation from circular motion became large. The median $v_{\rm col}$ became larger in all regions, especially the largest increase was seen in {\it arm}; $v_{\rm col} = 21.9~\rm km~s^{-1}$, $19.3~\rm km~s^{-1}$, and $16.6~\rm km~s^{-1}$ in {\it bar}, {\it bar-end}, and {\it arm}, respectively. The tendency of the $v_{\rm col}$ does not change, but the difference in $v_{\rm col}$ among the environments becomes small. This result suggests that our assumption results in underestimations of the collision frequency and velocity in {\it arm}. Note that we also assumed that GMCs are in constant linear motion in this check. Therefore, the paths of the pairs expected to collide with each other may change by the further collisionless encounters.



\begin{figure}
	\includegraphics[width=\hsize]{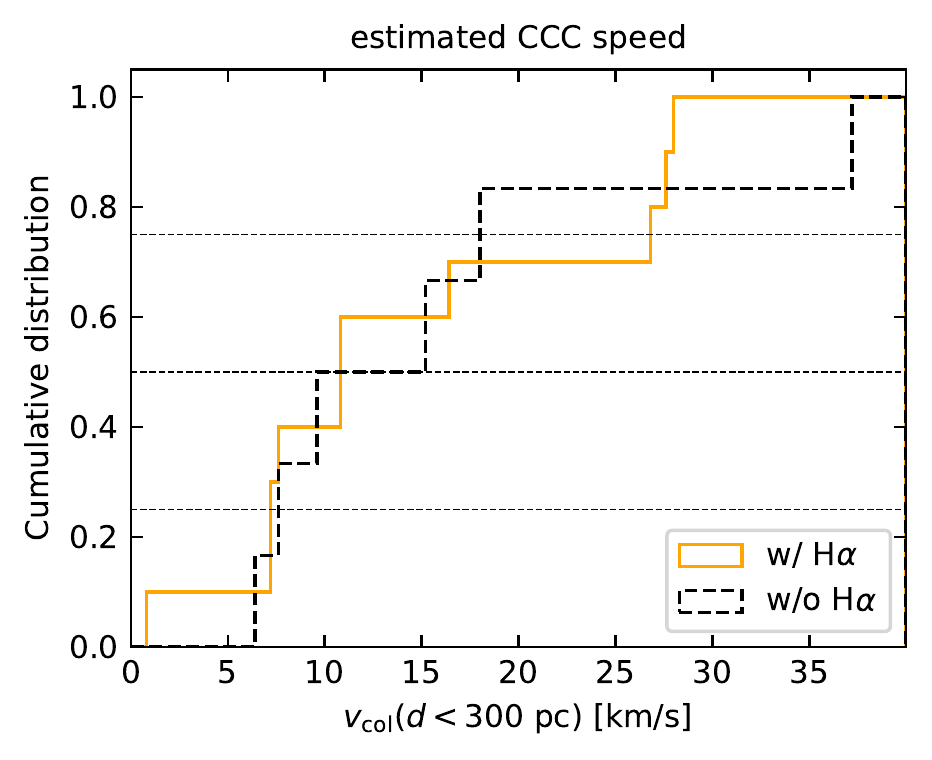}
    \caption{Normalized cumulative distribution function of the estimated collision frequency of GMCs with H$\alpha$ and without H$\alpha$. We use the GMCs in the {\it arm} and {\it bar-end}.}
 \label{fig:velocity_deviation_in_NGC1300_Ha}
\end{figure}

Forth assumption is that we implicitly neglect the impact of feedback on the GMCs. Local stellar feedback (e.g.  photoionisation, stellar winds, and supernovae) affects the GMC's formation and evolution \citep[e.g.][]{dobbs_properties_2011,tasker_star_2015,fujimoto_gmc_2016,grisdale_physical_2018}.
Recent studies show that
feedback mechanisms caused by massive star formation such as photoionisation and stellar winds play a major role in dispersing the GMC before the first supernova explosion \citep[e.g.][]{kruijssen_fast_2019, chevance_lifecycle_2020}. Therefore, a GMC  associated with H\textsc{ii} regions is likely to be affected by the feedback.
There may be no effect in the strong bar with no H\textsc{ii} regions, but we may have to consider the impact on the GMCs in {\it arm} and {\it bar-end} regions where the prominent H\textsc{ii} regions are seen. 
Fig.~\ref{fig:velocity_deviation_in_NGC1300_Ha}
shows the comparison of the CCC speeds of the pairs with H$\alpha$ and without H$\alpha$ in {\it arm} and {\it bar-end} regions.
Here, we used the continuum-subtracted H$\alpha$ image of NGC~1300 obtained from HST images \citep[see Fig.9 in][]{maeda_large_2020}. We don't see any difference between both distributions,  indicating the effect of the feedback would be small. 


It is also necessary to mention the timescale of the GMC dispersal by the stellar feedback. Recent observations show the feedback timescale, which is the time between the emergence of the first ionizing photons due to massive star formation and the eventual dispersal of the GMC, is $1-5$~Myr \citep[e.g.][]{hollyhead_studying_2015,kruijssen_fast_2019,hygate_uncertainty_2019,chevance_lifecycle_2020}. This implies that the GMCs associated with H\textsc{ii} regions at the current time are likely to be dispersed soon.
Thus, many of the pairs in {\it arm} and {\it bar-end} regions we identified as likely to collide within 10~Myr may disappear before they collide. 

\subsubsection{Caveats on the comparison between observation and simulation} \label{sec:Caveats on the comparison between observation and simulation}
Above sections, we simply compared the results of the observations and the simulation, but we must note that the data quality and method of molecular cloud identification are different. In the simulation by \citetalias{fujimoto_fast_2020}, the cloud mass ranges from $5 \times 10^3~\rm M_\odot$ to $5 \times 10^6~\rm M_\odot$, radius ranges from 5~pc to 15~pc. On the other hand, the mass and size of observed GMCs by \citet{maeda_properties_2020} are larger than those in the simulation; catalogued mass and radius range from $6 \times 10^4~\rm M_\odot$ to $1.0 \times 10^7~\rm M_\odot$ and from 8~pc to 100~pc, respectively. This is because the spatial resolution of our ALMA observations is $\sim$40~pc, which is much larger than the 2~pc resolution of the simulation, and the mass completeness limit of the observation is estimated to be $2 \times \rm 10^5~M_\odot$. Therefore, our observations miss the low-mass GMCs, would overestimate the radius, and would blend the small GMCs. Further, the methods of identification of GMCs are different. In the simulation, a cloud is defined as a connected dense region that is enclosed by a contour at a threshold number density of $400~\rm cm^{-3}$. This threshold would be higher than that in the observations  because the critical density of $^{12}$CO($1-0$) is lower than $400~\rm cm^{-3}$. Therefore, the simulation and observation might be looking at somewhat different objects. It is possible that a gas clump that is splited into multiple clouds in the simulation is identified as a single GMC in the observations. In order to make a more complete comparison, we need to match the spatial resolution between the simulation and observations by convolving the simulation with the beam size of the observations and identify the cloud with the same method. Further, similar to our analysis, the effect of GMC dispersal due to feedback on CCC is neglected in the simulation.
Therefore, the simulation should  include stellar feedback in the future.

\subsection{The cause for the different star formation activity between bar and bar-end with the same high CCC speed} \label{sec: The cause for the different star formation activity between bar and bar-end with the same high CCC speed}

Interestingly, the star formation activity in {\it bar} and {\it bar-end} regions, where the number density of high-speed CCC is similar, is quite different. The SFE is $0.08-0.12 \rm~Gyr^{-1}$ and $0.45 - 0.53 \rm~Gyr^{-1}$, respectively \citep{maeda_large_2020}. If the CCC is a triggering mechanism for massive star formation, our results show that star formation is triggered by high-speed CCC in {\it bar-end}, whereas it is not triggered in {\it bar}. This difference suggests that there are other parameters besides CCC speed that control the star formation.

One of the candidate parameters is the mass of the colliding molecular clouds. Previous studies  proposed the CCCs are responsible for the correlation between the SFE and the mass of the cloud
\citep[e.g.][]{scoville_high-mass_1986, ikuta_kinematical_1997, tan_star_2000}.
\citet{takahira_formation_2018} performed sub-parsec scale simulations of CCCs of two molecular clouds with different masses and with various collision speeds and investigated the formation of cores. They show the shape of the cumulative core mass function depends not only on the relative velocity of the clouds but also on the cloud mass (i.e. size). The number of massive cores tends to decrease with increasing collision speed and decreasing mass of the colliding  cloud. Therefore, a CCC with a higher relative velocity would require higher masses of the clouds to trigger star formation. Even if the collision velocity is the same, collisions between denser molecular clouds are more likely to trigger massive star formation.

In NGC~1300, the catalogued molecular gas mass of the GMC in {\it bar-end} is significantly larger that in {\it bar} \citep{maeda_properties_2020}. Further, a rough estimation of the average density of the GMC, $n_{\rm H_2}$, shows that $n_{\rm H_2}$ in {\it bar-end} ($\sim 70~\rm cm^{-3}$) is about two times larger than that in {\it bar} ($\sim 35~\rm cm^{-3}$). Here, $n_{\rm H_2}$ is defined as $M_{\rm mol}/(\frac{4}{3}\pi R^3)$. In fact, Fig. \ref{fig:Mmol_vs_Vcol} shows the molecular gas mass of the most massive GMC in each pair in {\it bar-end} is larger than that in  {\it bar}.
Thus, the difference in the molecular gas mass (density) would be cause for the different star formation activity between {\it bar} and {\it bar-end} with the same high CCC speed.
Recent CCC observations in the Milky Way are consistent with the conclusion of \citet{takahira_formation_2018}. Despite the same higher collision speed with $20 - 30~\rm km~s^{-1}$, CCCs with and without OB stars have been found; In the Sgr B2 and W 43, which are the famous star-burst regions in the Milky Way, CCCs with higher peak column density ($10^{23} - 10^{24}~\rm cm^{-2}$) are founded \citep{hasegawa_large-scale_1994,grisdale_physical_2018,kohno_forest_2020}. On the other hand, in the central region of the Milky Way, no OB stars are associated with a CCC with lower peak column density ($2 \times  10^{22}~\rm cm^{-2}$) in M-3.8+0.9 \citep{enokiya_cloud-cloud_2019}.

Prominent H\textsc{ii} regions in barred galaxies are preferentially located in the bar-end regions.
In NGC~1300, we see that the collision frequency of the GMCs is higher in {\it bar-end} than in {\it arm} based on the estimation of $v_{\rm col}$ and measurement of $\Delta_v$. This result suggests that active star formation is induced by frequent CCCs.
Previous observations and simulations towards the bar-end regions proposed the same scenario (e.g. M83; \citealt{kenney_orbit_1991}, NGC~3627; \citealt{beuther_interactions_2017}, simulations; \citealt{motte_formation_2014, renaud_environmental_2015}). They proposed that orbital crowding at the bar-end leads to CCC that can form massive molecular complexes and massive stellar associations. Recent observations by \citet{kohno_forest_2020} show that the high collision frequency in the W 43 GMC complex, which is located in the bar-end region of the Milky Way. They argue that converging gas flow from the arm and the bar causes highly turbulent condition leading to the high frequency.


\subsection{The cause of the low SFE in the bar region}\label{sec: The cause of the low SFE in the bar region}

As described in Section 1, massive star formation in bars of barred galaxies is suppressed. However, physical mechanisms of the suppression are still unclear. Several explanations have been proposed in observational and theoretical studies. Numerical hydrodynamic simulations of a strongly barred galaxy, NGC~5383, suggests that the molecular clouds are destroyed by shock due to the high velocity of the gases which flow into the bar structure \citep{tubbs_inhibition_1982}. Some hydrodynamic simulations and observations argued that existence of a high shear probably suppresses the molecular cloud formation \citep[e.g.][]{athanassoula_existence_1992,zurita_ionized_2004,emsellem_interplay_2015}. Observations towards NGC~1530 suggest a combination of the shock and the shear is the cause of the low star formation activity \citep{reynaud_kinematics_1998}.
It was proposed that gas in the bar regions is efficiently depleted due to the formation of the bar structure \citep[e.g.][]{spinoso_bar-driven_2017,james_star_2018}.

Recent other studies proposed  scenarios and physical mechanisms that could suppress the star formation fall into the following three scenarios: (1) A large amount of diffuse molecular gases in bar regions \citep[e.g.][]{muraoka_co_2016,yajima_co_2019, torii_forest_2019}. (2) Gravitationally unbound molecular clouds in bar regions \citep{sorai_properties_2012,nimori_dense_2013}. (3) High-speed CCCs in bar regions \citep{fujimoto_environmental_2014,fujimoto_fast_2020}.
Here, we discuss these scenarios based on previous observations of NGC~1300 \citep{maeda_properties_2020,maeda_large_2020} and the results of this work.

\citet{maeda_large_2020} tested scenario (1). Molecular gas traced by $^{12}$CO($1-0$) is supposed to consist of two components: GMCs and the extended diffuse component that is distributed on scales larger than sub-kpc \citep[see e.g. ][]{pety_plateau_2013, caldu-primo_spatially_2015,querejeta_alma_2020}.
In the SFE calculation using CO($1-0$) emission, the diffuse molecular gas, which would not directly contribute to the current star formation activity, is included in the $\Sigma_{\rm H_2}$. Thus, in other words, the diffuse gas greatly contributes to the $\Sigma_{\rm H_2}$ in the bar regions and that the SFE becomes low in appearance.
To examine the relationship between the SFE and the amount of the diffuse molecular gas, \citet{maeda_large_2020} measured the diffuse molecular gas fraction, $f_{\rm dif}$, in NGC~1300. Here, the $f_{\rm dif}$ is defined as $f_{\rm dif} = 1 - I_{\rm CO(1-0)}^{\rm ALMA}/I_{\rm CO(1-0)}^{\rm tot}$,
where $I_{\rm CO(1-0)}^{\rm ALMA}$ is CO($1-0$) flux  obtained from the ALMA 12-m array, which has no sensitivity on diffuse (full width at half-maximum $\gtrsim 700$~pc) molecular gases due to the lack of ACA, and $I_{\rm CO(1-0)}^{\rm tot}$ is the total CO($1-0$) flux obtained from the Nobeyama 45-m single-dish telescope. Thus, the diffuse molecular gas is defined as a molecular gas that is spread on a scale of 700~pc. They found  the environmental dependence of the $f_{\rm dif}$ and that the SFE decreases with increasing in $f_{\rm dif}$; The $f_{\rm dif}$ and ${\rm SFE}$ are $0.74 - 0.91$ and  $(0.06 - 0.16) ~\rm Gyr^{-1}$ in {\it bar}, and  $0.28 - 0.65$ and $(0.23 - 0.96) ~\rm Gyr^{-1}$ in {\it arm} and {\it bar-end}. This result supports scenario (1).
If the low SFE in the bar comes only from the large diffuse molecular gas fraction, recalculated SFEs using the molecular gas excluding the diffuse components (i.e. molecular gas organized inside GMCs) in the bar should be comparable to that in the arm and bar-end. However, the SFE using the molecular gas in GMCs is still significantly suppressed  in the bar regions \citep[see Fig. 17 in][]{maeda_large_2020}, which suggests the presence of other causes for the low SFE, besides a large amount of the diffuse molecular gases. There should be mechanisms that control the star formation process in the GMCs and scenarios (2) and (3) are the candidates.

\citet{maeda_properties_2020} investigated whether  scenario (2) is responsible for the suppression of SFE excluding the diffuse molecular gas in the bar region. Using a resolved GMC sample with $M_{\rm mol} > 5.0 \times 10^5~\rm M_\odot$ and $R > 15~\rm pc$, they investigated environmental dependence of the distribution of the virial parameter, which is a measure for gravitational binding of molecular clouds \citep{bertoldi_pressure-confined_1992}. There is no clear difference in the virial parameter between the GMCs in {\it bar} and those in {\it arm} and {\it bar-end}. This result suggests that the lack of massive star formation in the strong bar of NGC 1300  cannot be explained by a systematic difference in the virial parameter (i.e. scenario (2)). Note that the virial parameter of the GMCs with smaller masses ($< 5 \times 10^5~\rm M_\odot$) are unknown and additional observations would be required.

In this study, we investigated scenario (3) and find that the CCC speeds in {\it bar} and {\it bar-end} tend to be higher than that in {\it arm}. 
As discussed in Section~\ref{sec: The cause for the different star formation activity between bar and bar-end with the same high CCC speed}, not only the CCC speed but also the mass (density) of the colliding GMCs is a key parameter that controls the star formation activity of the GMCs. The lower molecular gas mass (density) would be the cause for the different star formation activity between {\it bar} and {\it bar-end} with the same high CCC speed. Compared {\it arm} and {\it bar}, the molecular gas mass of the colliding GMCs is comparable in both environments (Fig. \ref{fig:Mmol_vs_Vcol}). Therefore, the higher CCC speed would be the cause for the difference in the star formation activity between {\it arm} and {\it bar} with the same molecular gas mass. From this discussion, the cause for the low SFE of the GMCs in the bar regions of strongly barred galaxies would be the high-speed CCCs between lower mass (density) GMCs.


To summarize the discussion, the leading candidates of causes for the star formation suppression in the bar region in NGC~1300 are the presence of a large amount of diffuse molecular gases, which would not contribute to the star formation and makes the SFE low in appearance, and high-speed CCCs between lower mass (density) GMCs, which shorten the gas accretion phase of the cloud cores formed, leading to suppression of core growth and massive star formation.

\section{Summary}\label{sec:summary}

Recent simulations proposed that a CCC speed is different among  galactic-scale environments, which is responsible for the observed differences in SFE. In particular, it is proposed that a high-speed CCC can suppress  massive star formation in the bar regions. To examine this scenario, we investigated the CCC speed and its possible effect on the SFE in the strongly barred galaxy NC~1300, which is one of the suitable laboratories because the absence of the massive star formation is clearly seen in the bar regions. We used two methods. One is the estimation of the CCC speed  based on the mean velocity field of the molecular gas and catalogued line-of-sight velocity of the GMCs. The other is the measurement of the velocity deviation between the GMC and its surrounding GMCs, which is a model-independent observable and  a good indicator of the qualitative trends of the CCC speed. Our main results are as follows:

\begin{enumerate}
    \item We find environmental dependence in the estimated CCC speed. The speed tends to be higher in {\it bar} ($\sim20~\rm km~s^{-1}$) and {\it bar-end} ($\sim16~\rm km~s^{-1}$) than in {\it arm} ($\sim11~\rm km~s^{-1}$) (Section~\ref{sec:CCC Results}, Fig.~\ref{fig:collision_velocity_in_NGC1300}, and Table~\ref{tab:CCC speed}).
    
    \item The number density of high-speed CCC with $v_{\rm col} > 20~\rm km~s^{-1}$ is higher in {\it bar} and {\it bar-end} than in {\it arm}. The high number density of high-speed CCC in both {\it bar} and {\it bar-end}, where the star formation activity is significantly different, suggests that the existence of other  parameters that control triggered star formation (Section ~\ref{sec:CCC Results}, and Table~\ref{tab:CCC speed}).

    \item The collision frequency is estimated to be $10 -25~{\rm Gyr^{-1}}$ for all regions. We find that the collision frequencies in {\it bar} and {\it bar-end} are about two times higher than that in {\it arm}. The lack of star formation activity despite the high collision frequency in {\it bar} suggests that the star formation activity does not control only by the collision frequency (Section~\ref{sec:CCC Results} and Table~\ref{tab:CCC speed}). 
    
    \item We compare the estimated CCC speed with the simulation by \citet{fujimoto_fast_2020}, which presented a hydrodynamical simulation of a strongly barred galaxy, using a stellar potential model of NGC~1300. In {\it bar} and {\it arm}, the estimated CCC speeds are generally comparable to the simulated CCC speed. However, in {\it bar-end}, the estimated CCC speed  ($\sim20~\rm km~s^{-1}$) is higher than the simulated CCC ($\sim10~\rm km~s^{-1}$), which suggests the deviation from the circular motion may be higher in {\it bar-end} than that expected in the simulation (Section~\ref{sec: Comparison of CCC speed with the simulation}).
    
    \item We find that the velocity deviation in {\it bar} and {\it bar-end} are larger than that in {\it arm}, which is qualitatively consistent with the estimation of CCC speed (Section~\ref{sec: velocity deviation}, Fig.~\ref{fig:velocity_deviation_in_NGC1300}, and Table~\ref{tab:Velocity deviation}).
    
    \item One of the other parameters besides the CCC speed that control the star formation process by the CCC may be the mass (density) of the colliding molecular cloud. Significant difference in molecular gas mass (average density) between {\it bar} (lower mass and lower density) and {\it bar-end} (higher mass and higher density) may be cause for the different star formation activity despite the same high CCC speed (Fig.~\ref{fig:Mmol_vs_Vcol} and Section~\ref{sec: The cause for the different star formation activity between bar and bar-end with the same high CCC speed}).
    
    \item Considering the relationship among the molecular gas mass of GMCs, the CCC speed, and star formation activity, one of the causes for the star formation suppression in the bar of NGC~1300 would be the high-speed CCCs between lower mass (density) GMCs. Further, combining our previous observations of NGC~1300 \citep{maeda_properties_2020,maeda_large_2020}, we conclude that the leading candidates of causes for the suppression are the presence of a large amount of diffuse molecular gases and high-speed CCCs between low mass GMCs (Section~\ref{sec: The cause of the low SFE in the bar region}).

\end{enumerate}

Since the connection between the environment, the CCC speed, and the star formation activity would depend on the properties of the host galaxy (e.g. the presence or absence of bar structure, bar strength, stellar mass, the presence or absence of interaction with other galaxies, and so on), estimation of CCC speed in many nearby galaxies is important as the next step of this study. Of course, direct observations of the CCC are essential for accurate measurements of the CCC speed in nearby galaxies because the collision velocities in this study are predictive values based on the kinematics of the GMCs. Although direct observations of CCC in nearby galaxies have just begun \citep[e.g.][]{Finn_2019ApJ...874..120F,sano_alma_2020, tokuda_alma_2020, muraoka_alma_2020},  the environmental dependence of the CCC properties will become clearer as the number of samples increases.

Finally, it is worth emphasizing here that unveiling the cause for the massive star formation suppression is also very important to understand the star formation process in high redshift galaxies. In the high redshift universe, merging is frequent and is considered to be a very important process for the formation and evolution of galaxies. Cloud collisions must be mainly responsible for the star formation in such galaxies at high(est) redshifts, but we do not know well the actual physical mechanism and process to produce stars by cloud collision. Therefore,  understanding conditions of the CCC speed and the properties of colliding molecular clouds (e.g. mass, density) for star formation is important to obtain a better understanding of galaxy evolution.

\section*{Data Availability}
The data underlying this article are available in \url{http://almascience.nrao.edu/aq/}, and can be accessed with ALMA project ID: 2015.1.00925.S; 2017.1.00248.S.; and 2018.1.01651.S.
The derived data generated in this research will be shared on reasonable request to the corresponding author.

\section*{Acknowledgements}
We would like to thank the referee for useful comments and sugges- tions. 
We are grateful to F. Egusa for her comments in our analysis. FM is supported by Research Fellowship for Young Scientists from the Japan Society of the Promotion of Science (JSPS). KO is supported by JSPS KAKENHI Grant Numbers JP16K05294 and JP19K03928.
AH is funded by the JSPS KAKENHI Grant Number JP19K03923.
The Nobeyama 45-m radio telescope is operated by NRO, a branch of National Astronomical Observatory of Japan (NAOJ).
This paper makes use of the following ALMA data: ADS/JAO.ALMA\#2015.1.00925.S, \#2017.1.00248.S., and
\#2018.1.01651.S. ALMA is a partnership of ESO (representing its member states), NSF (USA) and NINS (Japan), together with NRC (Canada), MOST and ASIAA (Taiwan), and KASI (Republic of Korea), in cooperation with the Republic of Chile. The Joint ALMA Observatory is operated by ESO, AUI/NRAO and NAOJ.
Data analysis was in part carried out on the Multi-wavelength Data Analysis System operated by the Astronomy Data Center (ADC), NAOJ.

\bibliographystyle{mnras}
\bibliography{Reference_collision_velocity} 

\begin{thebibliography}{}
\makeatletter
\relax
\def\mn@urlcharsother{\let\do\@makeother \do\$\do\&\do\#\do\^\do\_\do\%\do\~}
\def\mn@doi{\begingroup\mn@urlcharsother \@ifnextchar [ {\mn@doi@}
  {\mn@doi@[]}}
\def\mn@doi@[#1]#2{\def\@tempa{#1}\ifx\@tempa\@empty \href
  {http://dx.doi.org/#2} {doi:#2}\else \href {http://dx.doi.org/#2} {#1}\fi
  \endgroup}
\def\mn@eprint#1#2{\mn@eprint@#1:#2::\@nil}
\def\mn@eprint@arXiv#1{\href {http://arxiv.org/abs/#1} {{\tt arXiv:#1}}}
\def\mn@eprint@dblp#1{\href {http://dblp.uni-trier.de/rec/bibtex/#1.xml}
  {dblp:#1}}
\def\mn@eprint@#1:#2:#3:#4\@nil{\def\@tempa {#1}\def\@tempb {#2}\def\@tempc
  {#3}\ifx \@tempc \@empty \let \@tempc \@tempb \let \@tempb \@tempa \fi \ifx
  \@tempb \@empty \def\@tempb {arXiv}\fi \@ifundefined
  {mn@eprint@\@tempb}{\@tempb:\@tempc}{\expandafter \expandafter \csname
  mn@eprint@\@tempb\endcsname \expandafter{\@tempc}}}

\bibitem[\protect\citeauthoryear{Aouad, James  \& Chilingarian}{Aouad
  et~al.}{2020}]{aouad_coupling_2020}
Aouad C.~J.,  James P.~A.,   Chilingarian I.~V.,  2020, \mn@doi [\mnras]
  {10.1093/mnras/staa1945}, 496, 5211

\bibitem[\protect\citeauthoryear{Athanassoula}{Athanassoula}{1992}]{athanassoula_existence_1992}
Athanassoula E.,  1992, \mn@doi [\mnras] {10.1093/mnras/259.2.345}, 259, 345

\bibitem[\protect\citeauthoryear{Bertoldi \& McKee}{Bertoldi \&
  McKee}{1992}]{bertoldi_pressure-confined_1992}
Bertoldi F.,  McKee C.~F.,  1992, \mn@doi [\apj] {10.1086/171638}, 395, 140

\bibitem[\protect\citeauthoryear{Beuther, Meidt, Schinnerer, Paladino  \&
  Leroy}{Beuther et~al.}{2017}]{beuther_interactions_2017}
Beuther H.,  Meidt S.,  Schinnerer E.,  Paladino R.,   Leroy A.,  2017, \mn@doi
  [\aap] {10.1051/0004-6361/201526749}, 597, A85

\bibitem[\protect\citeauthoryear{Bryan et~al.,}{Bryan
  et~al.}{2014}]{bryan_enzo_2014}
Bryan G.~L.,  et~al., 2014, \mn@doi [\apjs] {10.1088/0067-0049/211/2/19}, 211,
  19

\bibitem[\protect\citeauthoryear{Caldú-Primo, Schruba, Walter, Leroy, Bolatto
  \& Vogel}{Caldú-Primo et~al.}{2015}]{caldu-primo_spatially_2015}
Caldú-Primo A.,  Schruba A.,  Walter F.,  Leroy A.,  Bolatto A.~D.,   Vogel
  S.,  2015, \mn@doi [\aj] {10.1088/0004-6256/149/2/76}, 149, 76

\bibitem[\protect\citeauthoryear{Chevance et~al.,}{Chevance
  et~al.}{2020}]{chevance_lifecycle_2020}
Chevance M.,  et~al., 2020, \mn@doi [\mnras] {10.1093/mnras/stz3525}, 493, 2872

\bibitem[\protect\citeauthoryear{Colombo et~al.,}{Colombo
  et~al.}{2014}]{colombo_pdbi_2014}
Colombo D.,  et~al., 2014, \mn@doi [\apj] {10.1088/0004-637X/784/1/3}, 784, 3

\bibitem[\protect\citeauthoryear{{Cornwell}}{{Cornwell}}{2008}]{Cornwell2008}
{Cornwell} T.~J.,  2008, \mn@doi [IEEE Journal of Selected Topics in Signal
  Processing] {10.1109/JSTSP.2008.2006388}, \href
  {https://ui.adsabs.harvard.edu/abs/2008ISTSP...2..793C} {2, 793}

\bibitem[\protect\citeauthoryear{Dobbs, Burkert  \& Pringle}{Dobbs
  et~al.}{2011}]{dobbs_properties_2011}
Dobbs C.~L.,  Burkert A.,   Pringle J.~E.,  2011, \mn@doi [\mnras]
  {10.1111/j.1365-2966.2011.19346.x}, 417, 1318

\bibitem[\protect\citeauthoryear{Dobbs, Pringle  \& Duarte-Cabral}{Dobbs
  et~al.}{2015}]{dobbs_frequency_2015}
Dobbs C.~L.,  Pringle J.~E.,   Duarte-Cabral A.,  2015, \mn@doi [MNRAS]
  {10.1093/mnras/stu2319}, 446, 3608

\bibitem[\protect\citeauthoryear{Downes, Reynaud, Solomon  \& Radford}{Downes
  et~al.}{1996}]{downes_co_1996}
Downes D.,  Reynaud D.,  Solomon P.~M.,   Radford S. J.~E.,  1996, \mn@doi
  [ApJ] {10.1086/177046}, 461, 186

\bibitem[\protect\citeauthoryear{Emsellem, Renaud, Bournaud, Elmegreen, Combes
  \& Gabor}{Emsellem et~al.}{2015}]{emsellem_interplay_2015}
Emsellem E.,  Renaud F.,  Bournaud F.,  Elmegreen B.,  Combes F.,   Gabor
  J.~M.,  2015, \mn@doi [\mnras] {10.1093/mnras/stu2209}, 446, 2468

\bibitem[\protect\citeauthoryear{England}{England}{1989a}]{england_high-resolution_1989}
England M.~N.,  1989a, \mn@doi [\apj] {10.1086/167097}, 337, 191

\bibitem[\protect\citeauthoryear{England}{England}{1989b}]{england_dynamical_1989}
England M.~N.,  1989b, \mn@doi [\apj] {10.1086/167833}, 344, 669

\bibitem[\protect\citeauthoryear{Enokiya, Torii  \& Fukui}{Enokiya
  et~al.}{2019}]{enokiya_cloud-cloud_2019}
Enokiya R.,  Torii K.,   Fukui Y.,  2019, \mn@doi [\pasj]
  {10.1093/pasj/psz119}, p. psz119

\bibitem[\protect\citeauthoryear{{Finn}, {Johnson}, {Brogan}, {Wilson},
  {Indebetouw}, {Harris}, {Kamenetzky}  \& {Bemis}}{{Finn}
  et~al.}{2019}]{Finn_2019ApJ...874..120F}
{Finn} M.~K.,  {Johnson} K.~E.,  {Brogan} C.~L.,  {Wilson} C.~D.,  {Indebetouw}
  R.,  {Harris} W.~E.,  {Kamenetzky} J.,   {Bemis} A.,  2019, \mn@doi [\apj]
  {10.3847/1538-4357/ab0d1e}, \href
  {https://ui.adsabs.harvard.edu/abs/2019ApJ...874..120F} {874, 120}

\bibitem[\protect\citeauthoryear{Fujimoto, Tasker, Wakayama  \& Habe}{Fujimoto
  et~al.}{2014a}]{fujimoto_giant_2014}
Fujimoto Y.,  Tasker E.~J.,  Wakayama M.,   Habe A.,  2014a, \mn@doi [\mnras]
  {10.1093/mnras/stu014}, 439, 936

\bibitem[\protect\citeauthoryear{Fujimoto, Tasker  \& Habe}{Fujimoto
  et~al.}{2014b}]{fujimoto_environmental_2014}
Fujimoto Y.,  Tasker E.~J.,   Habe A.,  2014b, \mn@doi [\mnras]
  {10.1093/mnrasl/slu138}, 445, L65

\bibitem[\protect\citeauthoryear{Fujimoto, Bryan, Tasker, Habe  \&
  Simpson}{Fujimoto et~al.}{2016}]{fujimoto_gmc_2016}
Fujimoto Y.,  Bryan G.~L.,  Tasker E.~J.,  Habe A.,   Simpson C.~M.,  2016,
  \mn@doi [\mnras] {10.1093/mnras/stw1461}, 461, 1684

\bibitem[\protect\citeauthoryear{Fujimoto, Maeda, Habe  \& Ohta}{Fujimoto
  et~al.}{2020}]{fujimoto_fast_2020}
Fujimoto Y.,  Maeda F.,  Habe A.,   Ohta K.,  2020, \mn@doi [\mnras]
  {10.1093/mnras/staa840}, 494, 2131

\bibitem[\protect\citeauthoryear{{Fukui} et~al.,}{{Fukui}
  et~al.}{2014}]{Fukui2014ApJ...780}
{Fukui} Y.,  et~al., 2014, \mn@doi [\apj] {10.1088/0004-637X/780/1/36}, \href
  {https://ui.adsabs.harvard.edu/abs/2014ApJ...780...36F} {780, 36}

\bibitem[\protect\citeauthoryear{{Fukui} et~al.,}{{Fukui}
  et~al.}{2015}]{Fukui_2015ApJ...807L...4F}
{Fukui} Y.,  et~al., 2015, \mn@doi [\apjl] {10.1088/2041-8205/807/1/L4}, \href
  {https://ui.adsabs.harvard.edu/abs/2015ApJ...807L...4F} {807, L4}

\bibitem[\protect\citeauthoryear{Fukui, Habe, Inoue, Enokiya  \&
  Tachihara}{Fukui et~al.}{2020}]{fukui_cloud-cloud_2020}
Fukui Y.,  Habe A.,  Inoue T.,  Enokiya R.,   Tachihara K.,  2020,
  arXiv:2009.05077 [astro-ph]

\bibitem[\protect\citeauthoryear{Gilden}{Gilden}{1984}]{gilden_clump_1984}
Gilden D.~L.,  1984, \mn@doi [\apj] {10.1086/161894}, 279, 335

\bibitem[\protect\citeauthoryear{Grisdale, Agertz, Renaud  \& Romeo}{Grisdale
  et~al.}{2018}]{grisdale_physical_2018}
Grisdale K.,  Agertz O.,  Renaud F.,   Romeo A.~B.,  2018, \mn@doi [\mnras]
  {10.1093/mnras/sty1595}, 479, 3167

\bibitem[\protect\citeauthoryear{{Habe} \& {Ohta}}{{Habe} \&
  {Ohta}}{1992}]{Habe1992PASJ}
{Habe} A.,  {Ohta} K.,  1992, \pasj, \href
  {https://ui.adsabs.harvard.edu/abs/1992PASJ...44..203H} {44, 203}

\bibitem[\protect\citeauthoryear{Hakobyan et~al.,}{Hakobyan
  et~al.}{2016}]{hakobyan_supernovae_2016}
Hakobyan A.~A.,  et~al., 2016, \mn@doi [\mnras] {10.1093/mnras/stv2853}, 456,
  2848

\bibitem[\protect\citeauthoryear{Hasegawa, Sato, Whiteoak  \&
  Miyawaki}{Hasegawa et~al.}{1994}]{hasegawa_large-scale_1994}
Hasegawa T.,  Sato F.,  Whiteoak J.~B.,   Miyawaki R.,  1994, \mn@doi [\apjl]
  {10.1086/187417}, 429, L77

\bibitem[\protect\citeauthoryear{Hirota et~al.,}{Hirota
  et~al.}{2014}]{hirota_wide-field_2014}
Hirota A.,  et~al., 2014, \mn@doi [\pasj] {10.1093/pasj/psu006}, 66, 46

\bibitem[\protect\citeauthoryear{Hollyhead, Bastian, Adamo, Silva-Villa, Dale,
  Ryon  \& Gazak}{Hollyhead et~al.}{2015}]{hollyhead_studying_2015}
Hollyhead K.,  Bastian N.,  Adamo A.,  Silva-Villa E.,  Dale J.,  Ryon J.~E.,
  Gazak Z.,  2015, \mn@doi [\mnras] {10.1093/mnras/stv331}, 449, 1106

\bibitem[\protect\citeauthoryear{Hygate, Kruijssen, Chevance, Schruba, Haydon
  \& Longmore}{Hygate et~al.}{2019}]{hygate_uncertainty_2019}
Hygate A. P.~S.,  Kruijssen J. M.~D.,  Chevance M.,  Schruba A.,  Haydon D.~T.,
    Longmore S.~N.,  2019, \mn@doi [\mnras] {10.1093/mnras/stz1779}, 488, 2800

\bibitem[\protect\citeauthoryear{Ikuta \& Sofue}{Ikuta \&
  Sofue}{1997}]{ikuta_kinematical_1997}
Ikuta C.,  Sofue Y.,  1997, \mn@doi [PASJ] {10.1093/pasj/49.3.323}, 49, 323

\bibitem[\protect\citeauthoryear{James \& Percival}{James \&
  Percival}{2018}]{james_star_2018}
James P.~A.,  Percival S.~M.,  2018, \mn@doi [\mnras] {10.1093/mnras/stx2990},
  474, 3101

\bibitem[\protect\citeauthoryear{James, Bretherton  \& Knapen}{James
  et~al.}{2009}]{james_h_2009}
James P.~A.,  Bretherton C.~F.,   Knapen J.~H.,  2009, \mn@doi [\aap]
  {10.1051/0004-6361/200810715}, 501, 207

\bibitem[\protect\citeauthoryear{Kenney \& Lord}{Kenney \&
  Lord}{1991}]{kenney_orbit_1991}
Kenney J. D.~P.,  Lord S.~D.,  1991, \mn@doi [\apj] {10.1086/170634}, 381, 118

\bibitem[\protect\citeauthoryear{Kohno et~al.,}{Kohno
  et~al.}{2020}]{kohno_forest_2020}
Kohno M.,  et~al., 2020, \mn@doi [\pasj] {10.1093/pasj/psaa015}, p. psaa015

\bibitem[\protect\citeauthoryear{Kruijssen et~al.,}{Kruijssen
  et~al.}{2019}]{kruijssen_fast_2019}
Kruijssen J. M.~D.,  et~al., 2019, \mn@doi [Nature]
  {10.1038/s41586-019-1194-3}, 569, 519

\bibitem[\protect\citeauthoryear{Lang et~al.,}{Lang
  et~al.}{2020}]{lang_phangs_2020}
Lang P.,  et~al., 2020, \mn@doi [\apj] {10.3847/1538-4357/ab9953}, 897, 122

\bibitem[\protect\citeauthoryear{Leroy et~al.,}{Leroy
  et~al.}{2013}]{leroy_molecular_2013}
Leroy A.~K.,  et~al., 2013, \mn@doi [\aj] {10.1088/0004-6256/146/2/19}, 146, 19

\bibitem[\protect\citeauthoryear{Lindblad, Kristen, Joersaeter  \&
  Hoegbom}{Lindblad et~al.}{1997}]{lindblad_velocity_1997}
Lindblad P. A.~B.,  Kristen H.,  Joersaeter S.,   Hoegbom J.,  1997, \aap, 317,
  36

\bibitem[\protect\citeauthoryear{Loren}{Loren}{1976}]{loren_colliding_1976}
Loren R.~B.,  1976, \mn@doi [\apj] {10.1086/154741}, 209, 466

\bibitem[\protect\citeauthoryear{Maeda, Ohta, Fujimoto, Habe  \& Baba}{Maeda
  et~al.}{2018}]{maeda_large_2018}
Maeda F.,  Ohta K.,  Fujimoto Y.,  Habe A.,   Baba J.,  2018, \mn@doi [\pasj]
  {10.1093/pasj/psy028}, 70

\bibitem[\protect\citeauthoryear{Maeda, Ohta, Fujimoto  \& Habe}{Maeda
  et~al.}{2020a}]{maeda_properties_2020}
Maeda F.,  Ohta K.,  Fujimoto Y.,   Habe A.,  2020a, \mn@doi [\mnras]
  {10.1093/mnras/staa556}, 493, 5045

\bibitem[\protect\citeauthoryear{Maeda, Ohta, Fujimoto, Habe  \& Ushio}{Maeda
  et~al.}{2020b}]{maeda_large_2020}
Maeda F.,  Ohta K.,  Fujimoto Y.,  Habe A.,   Ushio K.,  2020b, \mn@doi
  [\mnras] {10.1093/mnras/staa1296}, 495, 3840

\bibitem[\protect\citeauthoryear{Momose, Okumura, Koda  \& Sawada}{Momose
  et~al.}{2010}]{momose_star_2010}
Momose R.,  Okumura S.~K.,  Koda J.,   Sawada T.,  2010, \mn@doi [\apj]
  {10.1088/0004-637X/721/1/383}, 721, 383

\bibitem[\protect\citeauthoryear{Motte et~al.,}{Motte
  et~al.}{2014}]{motte_formation_2014}
Motte F.,  et~al., 2014, \mn@doi [\aap] {10.1051/0004-6361/201323001}, 571, A32

\bibitem[\protect\citeauthoryear{{Mould} et~al.,}{{Mould}
  et~al.}{2000}]{MouldEtAl00}
{Mould} J.~R.,  et~al., 2000, \mn@doi [\apj] {10.1086/308304}, \href
  {https://ui.adsabs.harvard.edu/abs/2000ApJ...529..786M} {529, 786}

\bibitem[\protect\citeauthoryear{Muraoka et~al.,}{Muraoka
  et~al.}{2016}]{muraoka_co_2016}
Muraoka K.,  et~al., 2016, \mn@doi [\pasj] {10.1093/pasj/psw080}, 68, 89

\bibitem[\protect\citeauthoryear{Muraoka et~al.,}{Muraoka
  et~al.}{2020}]{muraoka_alma_2020}
Muraoka K.,  et~al., 2020, arXiv:2009.05804 [astro-ph]

\bibitem[\protect\citeauthoryear{Nimori, Habe, Sorai, Watanabe, Hirota  \&
  Namekata}{Nimori et~al.}{2013}]{nimori_dense_2013}
Nimori M.,  Habe A.,  Sorai K.,  Watanabe Y.,  Hirota A.,   Namekata D.,  2013,
  \mn@doi [\mnras] {10.1093/mnras/sts487}, 429, 2175

\bibitem[\protect\citeauthoryear{Pety et~al.,}{Pety
  et~al.}{2013}]{pety_plateau_2013}
Pety J.,  et~al., 2013, \mn@doi [\apj] {10.1088/0004-637X/779/1/43}, 779, 43

\bibitem[\protect\citeauthoryear{Querejeta et~al.,}{Querejeta
  et~al.}{2020}]{querejeta_alma_2020}
Querejeta M.,  et~al., 2020, arXiv:2011.01287 [astro-ph]

\bibitem[\protect\citeauthoryear{Randriamampandry, Combes, Carignan  \&
  Deg}{Randriamampandry et~al.}{2015}]{randriamampandry_estimating_2015}
Randriamampandry T.~H.,  Combes F.,  Carignan C.,   Deg N.,  2015, \mn@doi
  [\mnras] {10.1093/mnras/stv2147}, 454, 3743

\bibitem[\protect\citeauthoryear{Randriamampandry, Deg, Carignan  \&
  Widrow}{Randriamampandry et~al.}{2018}]{randriamampandry_simulating_2018}
Randriamampandry T.~H.,  Deg N.,  Carignan C.,   Widrow L.~M.,  2018, \mn@doi
  [\aap] {10.1051/0004-6361/201833509}, 618, A106

\bibitem[\protect\citeauthoryear{Renaud et~al.,}{Renaud
  et~al.}{2015}]{renaud_environmental_2015}
Renaud F.,  et~al., 2015, \mn@doi [\mnras] {10.1093/mnras/stv2223}, 454, 3299

\bibitem[\protect\citeauthoryear{Reynaud \& Downes}{Reynaud \&
  Downes}{1998}]{reynaud_kinematics_1998}
Reynaud D.,  Downes D.,  1998, \aap, 337, 671

\bibitem[\protect\citeauthoryear{{Rogstad}, {Lockhart}  \& {Wright}}{{Rogstad}
  et~al.}{1974}]{Rogstad_1974ApJ...193..309R}
{Rogstad} D.~H.,  {Lockhart} I.~A.,   {Wright} M.~C.~H.,  1974, \mn@doi [\apj]
  {10.1086/153164}, \href
  {https://ui.adsabs.harvard.edu/abs/1974ApJ...193..309R} {193, 309}

\bibitem[\protect\citeauthoryear{Rosolowsky \& Leroy}{Rosolowsky \&
  Leroy}{2006}]{rosolowsky_biasfree_2006}
Rosolowsky E.,  Leroy A.,  2006, \mn@doi [\pasp] {10.1086/502982}, 118, 590

\bibitem[\protect\citeauthoryear{{Saigo} et~al.,}{{Saigo}
  et~al.}{2017}]{Saigo_2017ApJ...835..108S}
{Saigo} K.,  et~al., 2017, \mn@doi [\apj] {10.3847/1538-4357/835/1/108}, \href
  {https://ui.adsabs.harvard.edu/abs/2017ApJ...835..108S} {835, 108}

\bibitem[\protect\citeauthoryear{{Sandage} \& {Tammann}}{{Sandage} \&
  {Tammann}}{1981}]{Sandae_Tammann}
{Sandage} A.,  {Tammann} G.~A.,  1981, {A Revised Shapley-Ames Catalog of
  Bright Galaxies}.
Carnegie Institution, Washington

\bibitem[\protect\citeauthoryear{Sano et~al.,}{Sano
  et~al.}{2020}]{sano_alma_2020}
Sano H.,  et~al., 2020, \mn@doi [\pasj] {10.1093/pasj/psaa045}, p. psaa045

\bibitem[\protect\citeauthoryear{Schmidt, Bigiel, Klessen  \& de Blok}{Schmidt
  et~al.}{2016}]{schmidt_radial_2016}
Schmidt T.~M.,  Bigiel F.,  Klessen R.~S.,   de Blok W. J.~G.,  2016, \mn@doi
  [\mnras] {10.1093/mnras/stw011}, 457, 2642

\bibitem[\protect\citeauthoryear{Scoville, Sanders  \& Clemens}{Scoville
  et~al.}{1986}]{scoville_high-mass_1986}
Scoville N.~Z.,  Sanders D.~B.,   Clemens D.~P.,  1986, \mn@doi [\apjl]
  {10.1086/184785}, 310, L77

\bibitem[\protect\citeauthoryear{Sellwood \& Sánchez}{Sellwood \&
  Sánchez}{2010}]{sellwood_quantifying_2010}
Sellwood J.~A.,  Sánchez R.~Z.,  2010, \mn@doi [\mnras]
  {10.1111/j.1365-2966.2010.16430.x}, 404, 1733

\bibitem[\protect\citeauthoryear{Solomon, Rivolo, Barrett  \& Yahil}{Solomon
  et~al.}{1987}]{solomon_mass_1987}
Solomon P.~M.,  Rivolo A.~R.,  Barrett J.,   Yahil A.,  1987, \mn@doi [\apj]
  {10.1086/165493}, 319, 730

\bibitem[\protect\citeauthoryear{Sorai et~al.,}{Sorai
  et~al.}{2012}]{sorai_properties_2012}
Sorai K.,  et~al., 2012, \mn@doi [\pasj] {10.1093/pasj/64.3.51}, 64, 51

\bibitem[\protect\citeauthoryear{Spekkens \& Sellwood}{Spekkens \&
  Sellwood}{2007}]{spekkens_modeling_2007}
Spekkens K.,  Sellwood J.~A.,  2007, \mn@doi [\apj] {10.1086/518471}, 664, 204

\bibitem[\protect\citeauthoryear{Spinoso, Bonoli, Dotti, Mayer, Madau  \&
  Bellovary}{Spinoso et~al.}{2017}]{spinoso_bar-driven_2017}
Spinoso D.,  Bonoli S.,  Dotti M.,  Mayer L.,  Madau P.,   Bellovary J.,  2017,
  \mn@doi [MNRAS] {10.1093/mnras/stw2934}, 465, 3729

\bibitem[\protect\citeauthoryear{Stark \& Brand}{Stark \&
  Brand}{1989}]{stark_kinematics_1989}
Stark A.~A.,  Brand J.,  1989, \mn@doi [ApJ] {10.1086/167334}, 339, 763

\bibitem[\protect\citeauthoryear{Stephens}{Stephens}{1970}]{Stephens:1970ic}
Stephens M.~A.,  1970, Journal of the Royal Statistical Society: Series B
  (Methodological), 32, 115

\bibitem[\protect\citeauthoryear{Stone}{Stone}{1970a}]{stone_collisions_1970-1}
Stone M.~E.,  1970a, \mn@doi [\apj] {10.1086/150309}, 159, 277

\bibitem[\protect\citeauthoryear{Stone}{Stone}{1970b}]{stone_collisions_1970}
Stone M.~E.,  1970b, \mn@doi [\apj] {10.1086/150310}, 159, 293

\bibitem[\protect\citeauthoryear{Takahira, Tasker  \& Habe}{Takahira
  et~al.}{2014}]{takahira_cloud-cloud_2014}
Takahira K.,  Tasker E.~J.,   Habe A.,  2014, \mn@doi [\apj]
  {10.1088/0004-637X/792/1/63}, 792, 63

\bibitem[\protect\citeauthoryear{Takahira, Shima, Habe  \& Tasker}{Takahira
  et~al.}{2018}]{takahira_formation_2018}
Takahira K.,  Shima K.,  Habe A.,   Tasker E.~J.,  2018, \mn@doi [\pasj]
  {10.1093/pasj/psy011}, 70

\bibitem[\protect\citeauthoryear{Tan}{Tan}{2000}]{tan_star_2000}
Tan J.~C.,  2000, \mn@doi [\apj] {10.1086/308905}, 536, 173

\bibitem[\protect\citeauthoryear{Tasker \& Tan}{Tasker \&
  Tan}{2009}]{tasker_star_2009}
Tasker E.~J.,  Tan J.~C.,  2009, \mn@doi [ApJ] {10.1088/0004-637X/700/1/358},
  700, 358

\bibitem[\protect\citeauthoryear{Tasker, Wadsley  \& Pudritz}{Tasker
  et~al.}{2015}]{tasker_star_2015}
Tasker E.~J.,  Wadsley J.,   Pudritz R.,  2015, \mn@doi [\apj]
  {10.1088/0004-637X/801/1/33}, 801, 33

\bibitem[\protect\citeauthoryear{Tokuda et~al.,}{Tokuda
  et~al.}{2020}]{tokuda_alma_2020}
Tokuda K.,  et~al., 2020, \mn@doi [\apj] {10.3847/1538-4357/ab8ad3}, 896, 36

\bibitem[\protect\citeauthoryear{Torii et~al.,}{Torii
  et~al.}{2019}]{torii_forest_2019}
Torii K.,  et~al., 2019, \mn@doi [\pasj] {10.1093/pasj/psz033}, 71, S2

\bibitem[\protect\citeauthoryear{Tubbs}{Tubbs}{1982}]{tubbs_inhibition_1982}
Tubbs A.~D.,  1982, \mn@doi [\apj] {10.1086/159846}, 255, 458

\bibitem[\protect\citeauthoryear{Yajima et~al.,}{Yajima
  et~al.}{2019}]{yajima_co_2019}
Yajima Y.,  et~al., 2019, \mn@doi [\pasj] {10.1093/pasj/psz022}, 71, S13

\bibitem[\protect\citeauthoryear{Zurita, Relaño, Beckman  \& Knapen}{Zurita
  et~al.}{2004}]{zurita_ionized_2004}
Zurita A.,  Relaño M.,  Beckman J.~E.,   Knapen J.~H.,  2004, \mn@doi [\aap]
  {10.1051/0004-6361:20031049}, 413, 73

\makeatother
\end{thebibliography}












\bsp	
\label{lastpage}
\end{document}